\documentclass[3p,sort&compress
]{elsarticle}

\makeatletter
\def\@author#1{\g@addto@macro\elsauthors{\normalsize%
    \def\baselinestretch{1}%
    \upshape\authorsep#1\unskip\textsuperscript{%
      \ifx\@fnmark\@empty\else\unskip\sep\@fnmark\let\sep=,\fi
      \ifx\@corref\@empty\else\unskip\sep\@corref\let\sep=,\fi
      }%
    \def\authorsep{\unskip,\space}%
    \global\let\@fnmark\@empty
    \global\let\@corref\@empty  
    \global\let\sep\@empty}%
    \@eadauthor={#1}
}

\usepackage[usenames,dvipsnames,pdftex]{xcolor}

\usepackage{acro} 
\usepackage{amsmath,amssymb,amsfonts,mathrsfs}
\usepackage{miller}
\usepackage{multirow}
\usepackage{subeqnarray}
\usepackage[hang]{caption}
\usepackage[labelfont=footnotesize,textfont=footnotesize]{subcaption}
\usepackage{tikz}
\usetikzlibrary{arrows,shapes,fadings,backgrounds,decorations.shapes,patterns}
\usepackage{verbatim}
\usepackage{bm}
\usepackage{diagbox} 
\usepackage{booktabs}
\usepackage[separate-uncertainty = true]{siunitx}
\usepackage[percent]{overpic}
\usepackage[title]{appendix}
\usepackage{graphicx}
\usepackage{algorithm2e}
\usepackage[pdftex,                                     
bookmarksnumbered=true,
colorlinks=true,%
linkcolor=Ref,
citecolor=Ref,
urlcolor=Ref,
pdfpagelabels=false,
plainpages = false,
]{hyperref}%
\usepackage{doi}
\usepackage{xcolor}

\usepackage[capitalise]{cleveref}                                   
\crefname{algocf}{alg.}{algs.}
\Crefname{algocf}{Algorithm}{Algorithms}

\newlength{\diagramsize}
\DeclareGraphicsExtensions{.pdf,.png}

\newcommand{\revision}[1]{#1}

\newcommand{\term}[1]{\textsc{#1}}
\newcommand{\ie}{\textit{i.e.}}
\newcommand{\eg}{\textit{e.g.}}

\newcommand{\kB}{\ensuremath{k_\text{B}}}

\newcommand{\norm}[2][]{\ensuremath{\left|\left|{#2}\right|\right|\if\relax\detokenize{#1}\relax\else _{#1}\fi}}
\newcommand{\transpose}[1]{\ensuremath{{#1}^{\text T}}}
\newcommand{\inverse}[1]{\ensuremath{{#1}^{-1}}}
\newcommand{\invtranspose}[1]{\ensuremath{{#1}^{\text{-T}}}}

\newcommand{\tnsrfour}[1]{\ensuremath{\mathbb{#1}}}
\newcommand{\tnsr}[1]{\ensuremath{\mathbf{#1}}}
\newcommand{\tnsrgreek}[1]{\ensuremath{\bm{#1}}}
\newcommand{\vctr}[1]{\ensuremath{\mathbf{#1}}}

\newcommand{\stiffness}{\tnsrfour C}

\newcommand{\sPK}{\ensuremath{\tnsr S}}
\newcommand{\F}[1][]{\ensuremath{\tnsr F^{#1}}}

\newcommand{\GL}{\ensuremath{\tnsr E_\text{e}}}
\newcommand{\Fp}[1][]{\ensuremath{\tnsr F_\text{p}^{#1}}}
\newcommand{\Fpdot}[1][]{\ensuremath{\dot{\tnsr F}_\text{p}^{#1}}}

\newcommand{\Fpinv}[1][]{\ensuremath{\inverse{\Fp}}}
\newcommand{\Fpinvtranspose}[1][]{\invtranspose{\Fp}}
\newcommand{\Fe}[1][]{\ensuremath{\tnsr F_\text{e}^{#1}}}
\newcommand{\FeT}{\ensuremath{\tnsr F_\text{e}^{\text{T}}}}
\newcommand{\Feinv}[1][]{\ensuremath{\inverse{\Fe}}}

\newcommand{\Lp}{\ensuremath{\tnsr L_\text{p}}}

\newcommand{\PK}{\term{Piola}--\term{Kirchhoff}}

\newcommand{\lnameref}[1]{%
	\bgroup
	\let\nmu\MakeLowercase
	\nameref{#1}\egroup}
\newcommand{\fnameref}[1]{%
	\bgroup
	\def\nmu{\let\nmu\MakeLowercase}%
	\nameref{#1}\egroup}

\newcommand{\nmu}{}

\begin{document}
\begin{frontmatter}

\title{On the interaction of precipitates and tensile twins in magnesium alloys}

\author[sjtu,mpie]{C.~Liu}
\author[mpie,manchester]{P.~Shanthraj\corref{cor1}}
\ead{pratheek.shanthraj@manchester.ac.uk}
\cortext[cor1]{Corresponding author}
\author[manchester]{J.D.~Robson}
\author[mpie]{M.~Diehl}
\author[sjtu]{S.~Dong\corref{cor1}}
\ead{sjtuds@sjtu.edu.cn}
\author[sjtu]{J.~Dong}
\author[sjtu]{W.~Ding}
\author[mpie]{D.~Raabe}
\address[sjtu]{National Engineering Research Center of Light Alloy Net Forming and State Key Laboratory of Metal Matrix Composite, School of Materials Science and Engineering, Shanghai Jiao Tong University, 200240 Shanghai, PR China}
\address[mpie]{Max-Planck-Institut f\"ur Eisenforschung GmbH, Max-Planck-Str.~1, 40237 D\"usseldorf, Germany}
\address[manchester]{School of Materials, University of Manchester, MSS Tower, Manchester M13 9PL, UK}
\journal{Acta Materialia}
\begin{abstract}
Although magnesium alloys deform extensively through shear strains and crystallographic re-orientations associated with the growth of twins, little is known about the strengthening mechanisms associated with this deformation mode.
A crystal plasticity based phase field model for twinning is employed in this work to study the strengthening mechanisms resulting from the interaction between twin growth and precipitates.
The full-field simulations reveal in great detail the pinning and de-pinning of a twin boundary at individual precipitates, resulting in a maximum resistance to twin growth when the precipitate is partially embedded in the twin.
Furthermore, statistically representative precipitate distributions are used to systematically investigate the influence of key microstructural parameters such as precipitate orientation, volume fraction, size, and aspect ratio on the resistance to twin growth.
The results indicate that the effective critical resolved shear stress (CRSS) for twin growth increases linearly with precipitate volume fraction and aspect ratio.
For a constant volume fraction of precipitates, reduction of the precipitate size below a critical level produces a strong increase in the CRSS due to the \term{Orowan}-like strengthening mechanism between the twin interface and precipitates.
Above this level the CRSS is size independent.
The results are quantitatively and qualitatively comparable with experimental measurements and predictions of mean-field strengthening models.
Based on the results, guidelines for the design of high strength magnesium alloys are discussed.
\end{abstract}


\end{frontmatter}

\section{Introduction}
\label{sec: introduction}
Wrought magnesium (Mg) and Mg alloys usually exhibit significant plastic anisotropy due to their strong crystallographic textures and the large differences in the critical resolved shear stress (CRSS) for the various deformation modes, \ie\ basal \hkl<a>, prismatic \hkl<a>, pyramidal \hkl<a+c> slip systems, and tensile twins \citep{reed1957deformation,tang2014formation,koike2003activity,molodov2014ductility,molodov2016role,yoshinaga1964nonbasal,agnew2003study,chapuis2011temperature,yoshinaga1963deformation}. 
The effects of grain size, texture, solutes, and precipitates on the threshold stress of dislocation slip and tensile twin deformation modes in Mg alloys have been studied extensively \citep{Robson2012,Agnew2013,Nie2013,tome2011multi,nave2004microstructures,brown2005internal,beyerlein2010statistical,al2008room,barnett2007twinning}.
Precipitation hardening is a promising mechanism to strengthen the overall material but also to tune the interaction with individual deformation modes. 
Therefore, it offers the potential to attenuate or accentuate the mechanical asymmetry of Mg alloys \citep{Jain2010,Robson2011}. 
For strongly textured Mg alloys, the relative increase of the CRSS for twin growth compared to that of prismatic slip due to  precipitation strengthening has been suggested to play an important role for the yield asymmetry \citep{Robson2012}.

The influence of the volume fraction, morphology, and habit plane of precipitates on the hardening of slip systems in Mg alloys has been well characterized and understood in terms of both experiments and theoretical analysis \citep{Nie2003,Nie2012,Wang2016}. 
In Mg alloys, common precipitate types are plates on the basal plane in Mg-Al alloy systems, rods with long axis being parallel to the c-axis of the matrix in Mg-Zn alloy series, and plates on the \hkl{10-10} prismatic planes in Mg-RE (rare earth) alloy systems \citep{Nie2012}.
Based on \emph{in situ} neutron diffraction experiments and elastoplastic self-consistent (EPSC) modeling, it was found that prismatic plate-shaped precipitates in Mg-RE alloys specifically harden basal slip, leading to an CRSS increase from \num{12} to \SI{37}{\mega \pascal}, \ie\ an enhancement of over \SI{200}{\percent} \citep{Agnew2013}. 
This effect is much less pronounced in the case of prismatic slip, where the CRSS value was found to increase from \num{78} to \SI{92}{\mega \pascal}, \ie\ by merely \SI{18}{\percent} \citep{Agnew2013}.
Compression tests conducted on an aged Mg-\num{5}\%Zn alloy including c-axis rod precipitates showed a moderate increase of the CRSS for basal slip of \SI{17}{\mega \pascal}, \ie\ \SI{33}{\percent} \citep{Wang2015}.
Moreover, it was reported that prismatic slip is also effectively impeded by c-axis rod precipitates in a Mg-\num{6}\%Zn alloy \citep{Jain2013}.
In AZ31 alloys with plate-shaped precipitates, the CRSS for basal slip was found to increase slightly by \SI{5}{\mega \pascal} (\SI{13}{\percent}) in comparison to a CRSS increase for prismatic slip of \SI{25}{\mega \pascal} (\SI{40}{\percent}) \citep{Stanford2012}.
The strengthening effect of precipitates with different morphologies in aged Mg alloys on the slip systems has been quantitatively predicted by the \term{Orowan} hardening model \citep{Ardell1985}, based on the bowing out of dislocations around shear-resistant precipitates.
In that context the inter-particle spacing has been identified as a critical microstructural parameter for controlling the hardening effect against dislocation slip \citep{Nie2003,Wang2016}.

The quantitative determination of the increase of the CRSS for twinning due to precipitates in Mg alloys has been studied by single crystal  experiments \citep{Wang2015,Liu2018} and polycrystalline experiments combined with crystal plasticity simulations \citep{Stanford2012,Agnew2013}.
Yet, there is considerable scatter in the measured magnitude of the increase of the CRSS for twinning from different studies even on the same alloy.
In the polycrystalline case, mean-field crystal plasticity models, \eg\ the visco-plastic self-consistent (VPSC) model \citep{Tome1991} or an averaged \term{Taylor} factor have been used to derive the CRSS for twinning from the global measured mechanical response.
The CRSS for twinning is determined by fitting the global stress--strain curves and texture evolution between experiments and simulations via modifying the CRSS values of various deformation modes.

The estimated CRSS increment for twinning due to basal plate precipitates in a rolled AZ31 alloy is \SI{31}{\mega \pascal} \citep{Stanford2012} and \SI{46}{\mega \pascal} \citep{Kada2016} based on \term{Taylor} factor approximation on the same alloy.
It is worth to note that the direct interaction between twins and precipitates and the inhomogeneous distribution of the local stress states during deformation can not be considered in both types of mean-field approaches, which might cause the significant scatter of the estimated CRSS for twinning in aged Mg alloys.
In Mg-Zn alloys containing c-axis rod precipitates exposed to different aging treatments, the increase of the CRSS for twinning has been derived as \SI{29}{\mega \pascal} \citep{Jain2015} and \SI{62}{\mega \pascal} \citep{Stanford2009} based on VPSC simulations.
Besides the above mentioned experiments conducted on polycrystals, micro-pillar compression experiments have been used to directly measure the increase of the CRSS due to the existence of precipitates \citep{Wang2015,Liu2018,wang2019quantification}, but so far with limited success. 

Prismatic plate precipitates usually form in Mg-RE alloys.
Due to the weak crystallographic texture of typical Mg-RE alloys, the important effect of solute RE on twinning, and the smaller relative differences among the CRSS values for the different slip modes in Mg-RE alloys, it is hard to reliably quantify the influence of prismatic plate precipitates on the threshold of twinning \citep{Agnew2013}.
Robson et al. \citep{Robson2016} and Fan et al. \citep{Fan2018} reported that the prismatic plate precipitates in Mg-RE alloys are effective obstacles against twin growth to reduce mechanical asymmetry, but on the contrary, it has been shown that the threshold for twinning activation in Mg-RE alloy after aging may be softened \citep{Agnew2013}. 
In the latter case, one possible reason is that solution depletion after aging may counterweight the strengthening effect of precipitates on twinning  \citep{Nie2013,Cui2017}.

From the pioneering work on the interaction between precipitates and twins in a Mg-\SI{8.5}{\percent}Al alloy by Gharghouri et al. \citep{Gharghouri1998,Gharghouri1999}, it is known that the size ratio between twins and precipitates determines whether twins impinge upon precipitates, bypass them by deviation of the twin habit plane, or fully engulf them.
Gharghouri et al. \citep{Gharghouri1998} also demonstrated that the large back-stress caused by the incompatibility strain between elastically deformed precipitates and the sheared (twinned) matrix needs to be relaxed by plastic deformation.
This is also evidenced by the observation of the concentration of dislocations ahead of precipitates in the matrix. 
Plastic relaxation around precipitates has also been confirmed in other Mg alloys with rod and basal plate-shaped precipitates  \citep{Stanford2009,Stanford2012}.
Furthermore, molecular dynamics (MD) simulations showed that basal dislocations or stacking faults nucleate from the twin precipitate interfaces when twin boundaries interact with large precipitates \citep{Fan2018}.
Clark \citep{Clark1965} reported that c-axis rod precipitates in a Mg-\SI{8.5}{\percent}Zn alloy are sheared by twins.
In contrast, recent studies on the same material showed that the precipitates are not sheared by the twin but exhibit a rigid rotation \citep{Stanford2009,Jain2015}. 

It is generally found that the volume fraction of tensile twins is either unaffected or slightly reduced by the presence of precipitates, however the average twin number density is significantly increased \citep{Gharghouri1998,Stanford2009,Jain2015}. 
This indicates that precipitates affect the CRSS for twin growth more than that for twin nucleation, which is rationalized on the basis that tensile twins preferably nucleate from grain boundaries within the precipitate free zone.
Furthermore, the inhibition of twin growth by precipitates increases the global applied stress and hence increases the probability of twin nucleation  \citep{Robson2010,Beyerlein2011}.

Although the different precipitates have been recognized as a significant factor for increasing the CRSS for twinning and thereby reducing the plastic anisotropy of Mg alloys, there is no consensus on the underlying strengthening mechanisms \citep{Robson2016,Fan2018}.
This lack of fundamental understanding makes it hard to reliably predict the influence of precipitate volume faction, size, morphology, and habit plane on the CRSS for twinning.
A common attempt to quantify the strengthening effect of precipitates on twinning is the use of the \term{Orowan} equation which was developed to quantify dislocation strengthening through particles \citep{Ardell1985,Nie2003}.
When using $\Delta \sigma \sim Gb/d$ for twins, a twinning partial dislocation can be treated in the same way as a normal, \ie\ non-dissociated slip dislocation, where $\Delta \sigma$ is the increase in strength, $G$ the shear modulus of the Mg matrix, $b$ the magnitude of the twinning partial dislocation and $d$ the effective planar inter-particle spacing on the twin plane \citep{Nie2003,Stanford2012}.
However, comparison to experimental results shows that the hardening predicted from this approach is substantially underestimated.
As the \term{Burgers} vector of the twinning partial dislocation, a critical factor determining the \term{Orowan} hardening, is approximately 5 times smaller than that for a slip dislocation, it can be deducted that additional effects play a role.

It has been proposed that the long-range back-stress---originating from the incompatibility strain between elastically deformed precipitates and the sheared matrix---is an important contribution to the inhibition of twin growth in addition to the \term{Orowan} stress  \citep{Jain2015,Robson2016}.
However, when using the \term{Eshelby} inclusion model approximation \citep{Brown1971,Brown1975}, the calculated back-stress---$\sim\gamma fG/2$, where $\gamma$ is the eigenstrain of twinning deformation and $f$ the volume fraction of precipitates---substantially overestimates the strengthening effects of precipitates on twin growth \citep{Jain2015,Robson2016}.
A possible explanation is the missing plastic deformation of the matrix in these calculations, although dislocations have been observed near precipitates in the twinned matrix in transmission electron microscopy (TEM) experiments and MD simulations \citep{Gharghouri1998,Fan2018}.
Moreover, precipitates are assumed to be completely rigid when calculating the back-stress, but it was observed that precipitates in Mg alloys are not very stiff and are thus able to deform elastically in response to the twinning induced shear deformation \citep{Stanford2009,Jain2015}. 
More recently, Fan et al. \citep{Fan2018} proposed a mean-field precipitation hardening model for twin growth considering precipitate size and shape effects, which is based on local energy conservation during the twin-precipitate interaction process.
The magnitude of the plastic relaxation energy to reduce the back-stress was approximately set to \SI{50}{\percent} of that in a pure elastic state.
However,  this value plays an important role on quantifying the strengthening effect against twinning and thus should be chosen carefully.

Due to the poor understanding of how twins interact with precipitates, strengthening models have identified different microstructural parameters as the critical factor to maximize the strengthening effect against twinning.
These parameters are the effective inter-particle spacing, $d$, in the \term{Orowan} hardening model \citep{Nie2003,Stanford2012} and the volume fraction of precipitates, $f$, in the back-stress model \citep{Jain2015,Robson2016}.
Therefore, a systematic study on the influence of precipitate volume fraction, morphology, orientation, complex stress distributions around precipitates as well as plastic relaxation on the twinning behavior is required for the design of high strength Mg alloys and the development of mechanism-based hardening models.

In the present work, the interactions between tensile twins and precipitates of different morphologies and habit planes are investigated at the meso-scale using an integrated full-field crystal plasticity(CP)-phase field(PF) model \citep{Liu_etal2018}.
The full-field CP-PF model bridges the gap between mean-field models, \eg\ the \term{Orowan} equation and the back-stress model mentioned above  \citep{Stanford2012,Jain2015,Robson2016}, and nano-scale approaches focusing on individual defects, \eg\ MD and Discrete Dislocation Dynamics (DDD).
It is inherently capable of describing the anisotropic elastic and plastic deformation resulting from dislocation slip-mediated and twinning induced deformation as well as predicting sources of back-stress, \eg\ the strain incompatibility between elastically deformed precipitates and the sheared matrix.
In particular, the influence of the precipitate orientation, precipitate volume fraction, precipitate size, and precipitate shape (aspect ratio) on the twin-precipitate interaction mechanisms was systematically investigated.
Their effect on the CRSS increment of twin growth was quantitatively determined and compared with experimental measurements as well as predictions of mean-field strengthening models.
Even though the simulations and analysis are performed for Mg alloys, the derived guidelines for effectively strengthening deformation twins by precipitation are expected to be applicable for other hexagonal close packed (HCP) alloys, \eg\ Ti and Zr alloys.

\section{Model formulation}
\label{sec: model}
The integrated full-field dislocation density-based CP and PF model for concurrently modeling of dislocation slip-mediated plasticity and heterogeneous twinning behavior including twin nucleation, propagation and growth in hexagonal metals in a finite strain framework was developed by Liu et al.  \citep{Liu_etal2018}. 
A spatially grain-resolved CP model is employed to predict the inhomogeneous distribution of stress, strain, crystallographic texture and dislocation activity.
In the present simulations a local CP model is applied and hence the additional back-stress due to dislocation pile-ups near precipitates, \ie\ a strain gradient effect \citep{FLECK1994475,Reuber_etal2014}, is not considered explicitly.
Twinning deformation in the PF model is driven by the \term{Ginzburg}-\term{Landau} relaxation of the total free energy including the orientation dependent twin interface energy and the elastic strain energy.
The model is briefly outlined in the following. For details readers are referred to Liu et al. \citep{Liu_etal2018}.  

\subsection{Phase field twinning model}
\label{sec: PhaseField}
The \term{Gibbs} free energy density of the considered system, $\psi$, includes the mechanical energy density, $\psi_{\mathrm{mech}}$, and the twin interface energy density, $\psi_{\mathrm{inf}}$, \ie\ 
\begin{equation}\label{eq: GibbsEnergy}
\begin{aligned}
\psi = \psi_{\mathrm{mech}} + \psi_{\mathrm{inf}}.
\end{aligned}      
\end{equation}
In the twin PF model, one non-conserved order parameter as a function of both time and position, $\varphi$, is used to describe the twinning process, \ie\ the trajectory along the minimum energy path (MEP) from the parent grain to the twinned crystal structure within the \term{Gibbs} free energy landscape \citep{Ishii2016}.
It equals to $\varphi=1$ inside the twin (twinned region) and $\varphi=0$ within the matrix (non-twinned region) and changes smoothly across the twin interface.
The twin interface energy density, $\psi_{\mathrm{inf}}(\nabla\varphi, \varphi)$, comprises the anisotropic gradient energy density and the crystalline energy density \citep{Hu20106554}: 
\begin{equation}\label{eq: TwnInfEnergy}
\begin{aligned}
\psi_{\mathrm{inf}}(\nabla\varphi, \varphi) = \tnsrgreek{\kappa} \cdot (\nabla \varphi \otimes \nabla \varphi) + \Delta f (1 - \varphi)\varphi,
\end{aligned}      
\end{equation}
where $\tnsrgreek{\kappa}$ is a symmetric second-order tensor related to the anisotropic twin interface energy.
Its components $\kappa_{11}, \kappa_{22}, \kappa_{33}$ represent the twin interface energy density values with respect to the twin tip, the lateral twin interface and the coherent twin boundary, respectively.
The crystalline energy, $\Delta f (1 - \varphi)\varphi$, is the potential energy landscape when an original parent crystal is sheared into a twin crystal orientation. 
$\Delta f$ is the energy barrier of the crystalline energy, which can be obtained by \emph{ab initio} calculations \citep{Ishii2016}.
The minima of the crystalline energy at $\varphi = \num{0.0}$ and $\varphi = \num{1.0}$ describe the equilibrium states of the parent and twin crystal, respectively.

The temporal evolution of the deformation twin is described by the time-dependent \term{Ginzburg}--\term{Landau} equation
\begin{equation}\label{eq:GinzBurg}
\begin{aligned}
\frac{\dot{\varphi}}{M} = \text{Div} ~ \tnsrgreek{\kappa} \nabla \varphi + \Delta f(2\varphi - 1)  - \partial_{\varphi} \psi_{\mathrm{mech}},
\end{aligned}      
\end{equation}
where M is the mobility parameter related to the twin boundary migration mobility, and $\psi_{\mathrm{mech}}$ is the mechanical contribution to the free energy density described in detail in \cref{sec: CrystalPlasticity}. 

\subsection{Crystal plasticity model}
\label{sec: CrystalPlasticity}
The total deformation gradient, \F, in a material point is multiplicatively decomposed as $\F = \Fe \Fp$.
An anisotropic elastic stiffness $\stiffness$ relates the elastic deformation gradient $\Fe$ to the second \PK\ stress by $\sPK = \stiffness \GL$.
\GL\ is the elastic \term{Green}--\term{Lagrange} strain, \ie\ $\GL=(\FeT \Fe - \tnsr{I})/2$.
\Fp\ is the plastic deformation gradient.
The rate evolution of the plastic deformation gradient, \ie\ $\Fpdot = \Lp \Fp $, is determined by the plastic velocity gradient, \Lp.
The plastic velocity gradient, \Lp, is additively composed from the plastic contributions from dislocation slip in the parent grain, shear from deformation twinning and subsequent dislocation slip in the twinned domain:

\begin{equation}\label{eq:Lp}
\begin{aligned}
\Lp = \bigg(1 - h_{\text{tw}}(\varphi) \bigg) \sum_{\alpha = 1}^{N_\text{s}} \dot{\gamma}^\alpha_\text{s} \vctr{m}^{\alpha}_\text{s} \otimes \vctr{n}^{\alpha}_\text{s} 
+ h_{\text{tw}}(\varphi) \sum_{\alpha = 1}^{N_{\text{s-}\text{tw}}} \dot{\gamma}^\alpha_{\text{s-}\text{tw}} \vctr{m}^{\alpha}_{\text{s-}\text{tw}} \otimes \vctr{n}^{\alpha}_{\text{s-}\text{tw}}
+ \dot{\gamma}_{\text{tw}} \vctr{m}_{\text{tw}} \otimes \vctr{n}_{\text{tw}},
\end{aligned}      
\end{equation}
in which $\dot{\gamma}^\alpha_s$  (see \cref{App: Plastic flow relations}) is the shear rate on the slip system $\alpha$,
and the vectors $\vctr{m}^{\alpha}_\text{s}$ and $\vctr{n}^{\alpha}_\text{s}$ indicate the slip direction and the slip plane normal of the $N_\text{s}$ slip systems in the parent crystal, respectively. 
Slip in the twinned region is denoted by the subscript ``s-tw''.
The vectors $\vctr{m}_{\text{tw}}$ and $\vctr{n}_{\text{tw}}$ indicate the twin direction and the twin plane normal of the twinning system considered.
$h_{\text{tw}}(\varphi) = 3 \varphi^2 - 2\varphi^3$ is a monotonically increasing interpolation function that describes the twin volume fraction within a material point.
Furthermore, $\dot{\gamma}_{\text{tw}} = \gamma_\text{ct} \partial_\varphi h_{\text{tw}}(\varphi) \dot{\varphi}$ is the shear rate on the twin system in which $\gamma_\text{ct}$ depicts the characteristic shear strain and has a value of \num{0.129} for the considered \hkl{-1012} tensile twin systems in Mg.
In this work, the basal \hkl<a> (\hkl{0001} \hkl<11-20>), prismatic \hkl<a> (\hkl{10-10} \hkl<11-20>) and pyramidal \hkl<c+a> (\hkl{-1-122} \hkl<-1-123>) slip systems are considered. 

A quadratic representation of the mechanical contribution, \ie\ the elastic strain energy density, is adopted:
\begin{equation}\label{eq:elast}
\psi_{\mathrm{mech}} = \frac{1}{2} \sPK \cdot \GL.
\end{equation}

The driving force for twin growth in the \term{Ginzburg}--\term{Landau} equation is then obtained as $\partial_{\varphi} \psi_{\mathrm{mech}} = - h_\text{tw}'(\varphi) \gamma_{ct} \sPK \cdot (\vctr{m}_\text{tw} \otimes \vctr{n}_\text{tw})$.
It can been seen that the mechanical driving force for twinning is the resolved shear stress on the twin plane along the twinning shear direction.

The coupled CP and PF model presented above is implemented within the D\"usseldorf Advanced Material Simulation Kit, DAMASK \citep{Roters_etal2012,Roters_etal2018}.
The solution of the equilibrium state of the coupled fields subjected to applied boundary conditions is solved by an in-house developed parallel finite element code \citep{Shanthraj_etal2016,Shanthraj_etal2017}.

\subsection{Simulation setup}
\label{sec: simulation_setup}
The simulation setup was designed to investigate the influence of precipitates on the growth of pre-existing tensile twins in Mg alloys.
To this end, a single crystal with a pre-inserted twin band was used as the starting configuration to isolate the twin-precipitate interaction and separate it from any other effects associated with twin-twin interactions and grain boundaries.
The precipitates with different orientation, volume fraction, size, and shape (aspect ratio) were randomly distributed in the single crystal.
The specific configurations are given in detail in the following section together with the results.
To obtain statistical characteristics of the calculation, five groups of randomly distributed precipitates were constructed for each case.
In the simulations, dislocation slip-mediated plasticity and twinning deformation are allowed to act as deformation carriers in the matrix, while the precipitates can only deform elastically.
This type of simulation set-up is based on results from experiments and MD simulations which show that the precipitates in Mg alloys are not sheared by twins \citep{Stanford2009,Jain2015,Fan2018}.
While it is known that the elastic stiffness of the precipitates depends on their compositions and lattice structures, here it is assumed to be the same as that of the matrix in order to focus on a general understanding of the interaction of precipitates and tensile twins.

The two dimensional (2D) single crystal was discretized into $512\times512$ (X and Z axis) regular hexahedral elements and the initial width of the pre-inserted twin band was taken as \num{10} elements.
A dimensionless length-scale was used in these simulations.
In-plane simple shear loading under periodic boundary conditions was imposed along the X direction to a total strain of \SI{11.0}{\percent} at a strain rate of \num{0.001} $\text{s}^{-1}$.
Note that the length scale along the thickness direction of the two dimensional model is infinite due to the periodic boundary condition, and hence precipitates extent to infinity along the normal direction of the 2D model (Y axis). 
The crystal orientation is given in \cref{fig: mechanicalResponses_singlePre}a with Euler angles of (\SI{90}{\degree}, \SI{-43}{\degree}, \SI{-60}{\degree}) in the \term{Bunge}-\term{Euler} angle notation, which means that the twinning direction is parallel to the applied shear direction.

\section{Results}
\label{sec: results}
\subsection{Interaction of plate-shaped precipitates with tensile twins}
\label{sec: singlePrecipitate}
\begin{figure}
	\centering
	\includegraphics[width=0.95\textwidth]{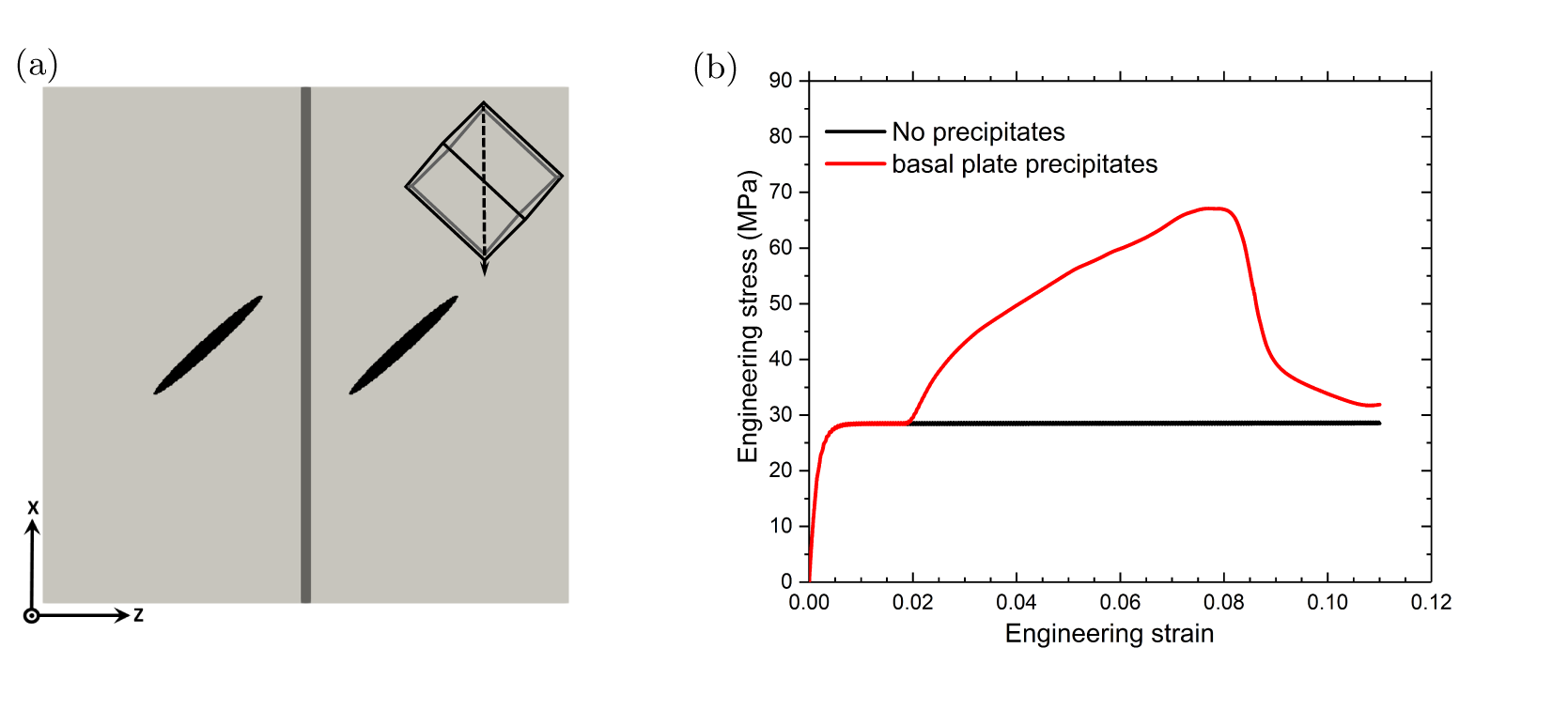}
	\caption{(a) The single crystal set-up containing two symmetrically placed basal plate precipitates (given in black color) and a pre-inserted twin band (given in dark grey color). The orientation of the hexagonal unit cell is shown in which the black arrow depicts the twinning direction of the active \hkl{-1012} twin.
		(b) The simulated stress--strain curves of the single crystal with and without precipitates.
	}
	\label{fig: mechanicalResponses_singlePre}
\end{figure}

Plate-shaped precipitates are generally expected to be effective strengtheners against tensile twin growth since they produce high incompatibility stress and remain unsheared in the twinned region.
$\beta-\text{Mg}_{17}\text{Al}_{12}$ basal plate precipitates are commonly observed in Mg-Al alloy systems and have a body-centered-cubic (BCC) structure.
Their orientation relationship with the $\alpha$-matrix is $\hkl(011)_\beta // \hkl(0001)_\alpha, \hkl[1-11]_\beta//\hkl[2-1-10]_\alpha$.
$\beta_1-\text{Mg}_3(\text{Gd,Nd,Y})$ prismatic plate precipitates are usually observed in Mg-RE alloys and have a face-centered-cubic (FCC) structure.
Their orientation relationship with the $\alpha$-matrix is $\hkl(-112)_{\beta_1}//\hkl(1-100)_\alpha$, $\hkl[110]_{\beta_1}//\hkl[0001]_\alpha$ \citep{Nie2012}.
In order to characterize the interaction process of tensile twins with plate-shaped precipitates in Mg alloys in detail, the interaction of an individual twin with isolated plate-shaped precipitates was simulated first.
Two basal plate precipitates were symmetrically placed close to the twin boundaries, as schematically shown in \cref{fig: mechanicalResponses_singlePre}a.
The aspect ratio of the basal plate precipitates was chosen to be 10 as generally observed in experiments \citep{Nie2012}.

\cref{fig: mechanicalResponses_singlePre}b shows the simulated stress--strain curves of the single crystal with and without the basal plate precipitates under shear deformation.
The threshold for twin growth in the absence of precipitates is around \SI{29}{\mega \pascal} corresponding to the horizontal plateau of the stress--strain curve.
The applied shear stress increases when twin boundaries approach the precipitates and reaches the peak value of \SI{67}{\mega \pascal}, then decreases rapidly once the twin has engulfed the precipitates.
It can be seen that the threshold stress for twin growth is still higher (around \SI{6}{\mega \pascal}) in the presence of precipitates even after the twin has fully engulfed the precipitates (corresponding to the applied shear strain from $\sim$ \SI{9.5}{\percent} to $\sim$ \SI{11}{\percent}).
This might be ascribed to the redistribution of the stress field and profuse dislocation generation in the matrix due to the twin-precipitate interaction.

\begin{figure}
	\centering
	\includegraphics[width=0.75\textwidth]{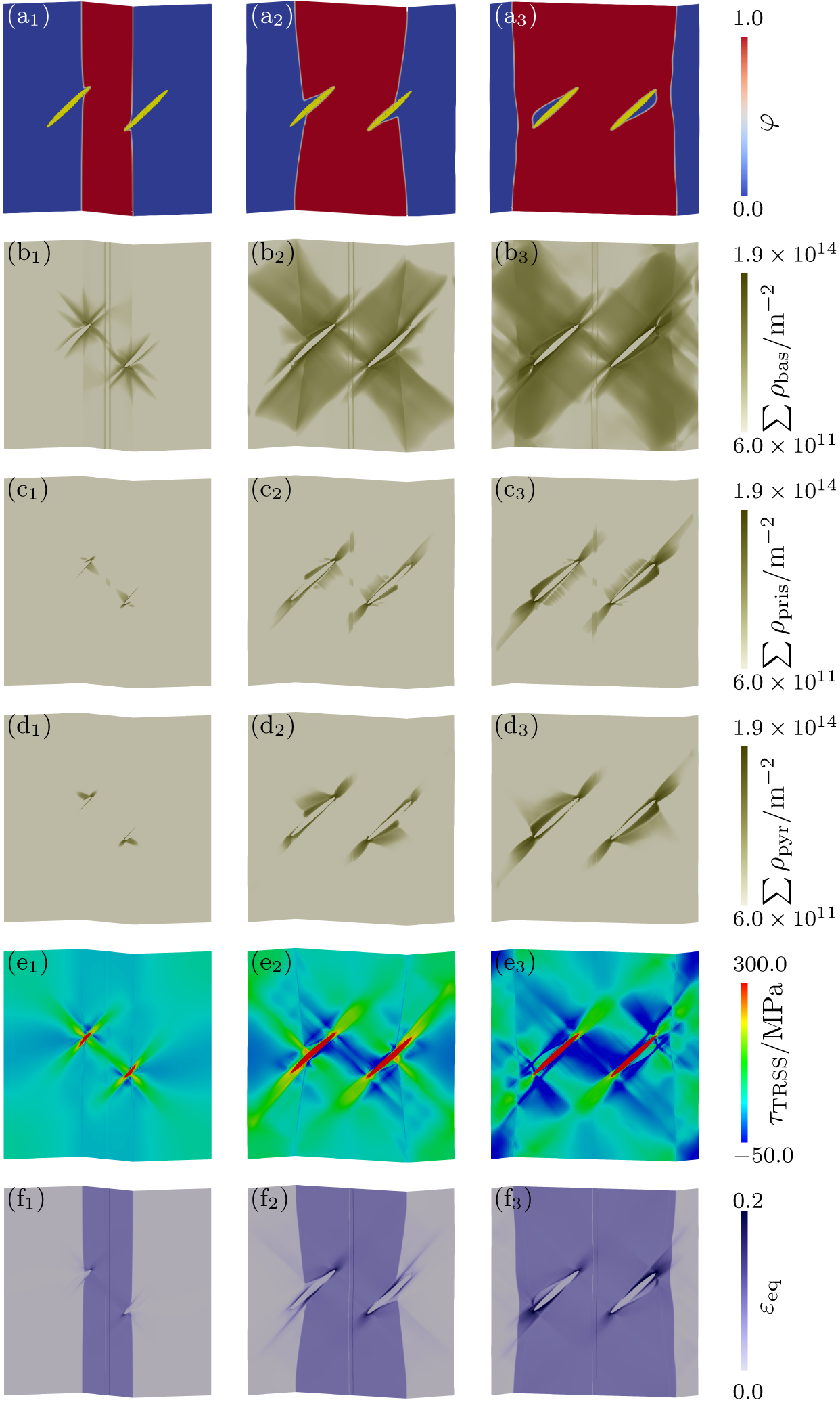}
	\caption{%
		Simulation results during the process of the interaction between a tensile twin and basal plate precipitates at different levels of shear deformation (left to right: \SI{3.3}{\percent}, \SI{6.7}{\percent}, \SI{10.0}{\percent}). Evolution of the (a) twinning microstructure where the precipitate is given in yellow color. The legend symbol $\varphi$ refers to the structure variable, \ie\ $\varphi=0$ within the matrix and $\varphi=1$ within the twin. (b) dislocation density of the basal slip system ($\rho_{\text{bas}}$), (c) dislocation density of the prismatic slip system ($\rho_{\text{pris}}$), (d) dislocation density of the pyramidal slip system ($\rho_{\text{pyr}}$), (e) resolved shear stress on the \hkl{-1012} twin system ($\tau_\text{TRSS}$) and (f) \term{von Mises} strain ($\varepsilon_\text{eq}$). 
		Note the logarithmic scale is used for all the dislocation density maps.
	}
	\label{fig: evolutionSiglePre}
\end{figure}

\cref{fig: evolutionSiglePre} presents the evolution of the twinning microstructure, activities of basal, prismatic, and pyramidal dislocations, resolved shear stress on the \hkl{-1012} twin system (TRSS) and \term{von Mises} strain distribution during the interaction between a tensile twin and basal plate precipitates at different stages of shear deformation.
It can be seen that the straight twin boundaries migrate towards the precipitates under the increasing applied strain and become pinned when they approach the particles.
Once the twin boundaries interact with the particles, the twin boundaries become arc-shaped. 
Even after the twin has completely engulfed the precipitates, the twin boundaries still exhibit a deflected shape and does not recover to the initial straight shape, as shown in \cref{fig: evolutionSiglePre}$\text{a}_3$.
It is worth to note that the twin boundaries migrate close to one side of the precipitates (top side for the right precipitate/bottom side for the left precipitate in \cref{fig: evolutionSiglePre}a) during the whole interaction process.
However, on the other side  of the precipitates, twin boundaries deviate from the precipitate surface after the twin has passed the precipitate tip.
The narrow region of the untwinned matrix between precipitates and twin boundaries contains a high dislocation density and even pyramidal slip systems with a high threshold stress are activated, see \cref{fig: evolutionSiglePre}(b-d).
These dislocations are required to accommodate the strain incompatibility between the elastically deformed precipitate and the sheared matrix.

\cref{fig: evolutionSiglePre}(b-d) show that both \hkl<a> and \hkl<a+c> slip systems are activated during the interaction process to relax the misfit strain which has both \hkl<a> and \hkl<c> components, however, the plastic relaxation zone accommodated by the basal slip is much larger than that by prismatic and pyramidal slip.
The non-basal slip activation is restricted to a region ahead of precipitates.
The predicted plastic relaxation region near precipitates is consistent with the limited experimental observations available.
However, the specific dislocation types are not able to be determined in these experiments due to a high dislocation density.
The current simulations show that extensive non-basal dislocations in addition to the ``easy'' basal slip are triggered to relieve the strain incompatibility during the twin-precipitate interaction process.

\cref{fig: evolutionSiglePre}e reveals that the stress distribution is extremely inhomogeneous due to the interaction between the twin band and the basal plate precipitates.
Stress significantly concentrates ahead of the precipitates and near the twin-precipitate intersection region which results in the activation of prismatic and pyramidal slip with a high CRSS value.
The stress far from the transverse precipitate habit plane is substantially relaxed.
Furthermore, it is predicted that the maximum stress state occurs when part of the precipitate is embedded in the twin whilst part is anchored in the matrix, see \cref{fig: evolutionSiglePre}$\text{e}_2$. 
It corresponds to the largest strain incompatibility state.
It can be seen from \cref{fig: evolutionSiglePre}f that the equivalent strain strongly concentrates in the vicinity of the precipitate tips, where high stress concentration and profuse accumulation of different types of dislocations were observed.

A direct determination of the strengthening effect of  prismatic plate precipitates for twin growth in Mg-RE alloys  based on experiments remains a hard challenge, since the weak crystallographic texture and solute RE have a profound influence on twinning.
Circumventing these effects, a similar simulation setup as presented above was therefore used to investigate the influence of prismatic plate precipitates on twin growth (not shown here). 
In the single crystal two prismatic plate precipitates were symmetrically placed close to the twin boundaries.
It is found that the twinning microstructure evolution, dislocation activities, and the mechanical response during the interaction process are similar to that of basal plate precipitates.
This can be attributed to the similar alignment angle between the twin plane and the precipitate habit plane, \ie\ \SI{43}{\degree} and \SI{47}{\degree} for basal plate and prismatic plate precipitates, respectively.

\subsection{Influence of precipitate habit plane}
\label{sec: habitPlane}
\begin{figure}
	\centering
	\includegraphics[width=0.8\textwidth]{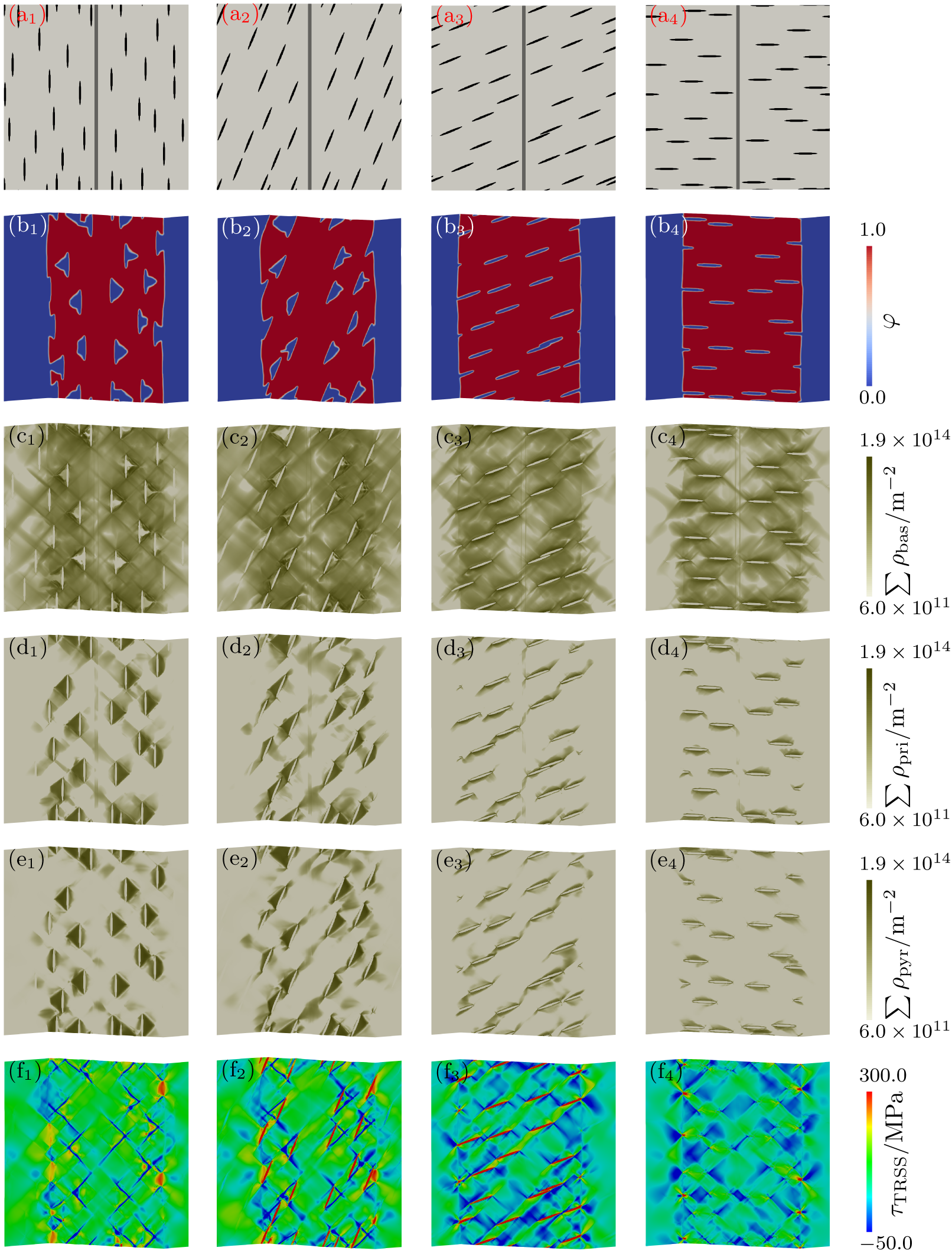}
	\caption{(a) Representative stress--strain curves of the single crystal with various precipitate habit plane. Plate-shaped precipitates with a total volume fraction of \SI{4}{\percent} and an aspect ratio of 10 are randomly distributed in the single crystal. $\theta$ is the alignment angle between the twin plane and the precipitate habit plane. $d$ is the precipitate diameter. 
		(b) Evolution of the CRSS increment ($\Delta \tau_\text{CRSS}$), average basal ($\rho_{\text{basal}}$), prismatic ($\rho_{\text{prismatic}}$), pyramidal ($\rho_{\text{pyramidal}}$) dislocation density in the twinned region, and back-stress ($\sigma_\text{back-stress}$) at a shear strain of \SI{8.3}{\percent} versus the alignment angle between the twin plane and precipitate habit plane. Note that the counterparts in the case containing spherical precipitates with a diameter of 20 elements and the same volume fraction of \SI{4}{\percent} is also given (d=20, \SI{4}{\percent}).
	}
	\label{fig: orientation}
\end{figure}

Similar to the orientation-dependent hardening effect of precipitates on dislocation slip, the orientation of the habit plane of plate-shaped precipitates is also expected to play an important role on the strengthening effect against  twin growth.
Here, simulations of the interaction between a tensile twin and multiple plate-shaped precipitates with different habit planes have been performed and the orientation-dependent strengthening effect was quantitatively determined.
The above introduced single crystal model with a single pre-existing twin band was used, in which multiple plate-shaped precipitates were randomly distributed (\cref{fig: orientation}a).
These plate-shaped precipitates have a volume fraction of \SI{4}{\percent} and an aspect ratio of \num{10}.
The angle between the twin plane and the precipitate habit plane, $\theta$, is varied from \SI{0}{\degree} to \SI{90}{\degree} with an increment of \SI{22.5}{\degree}.

\begin{figure}
	\centering
	\includegraphics[width=1.0\textwidth]{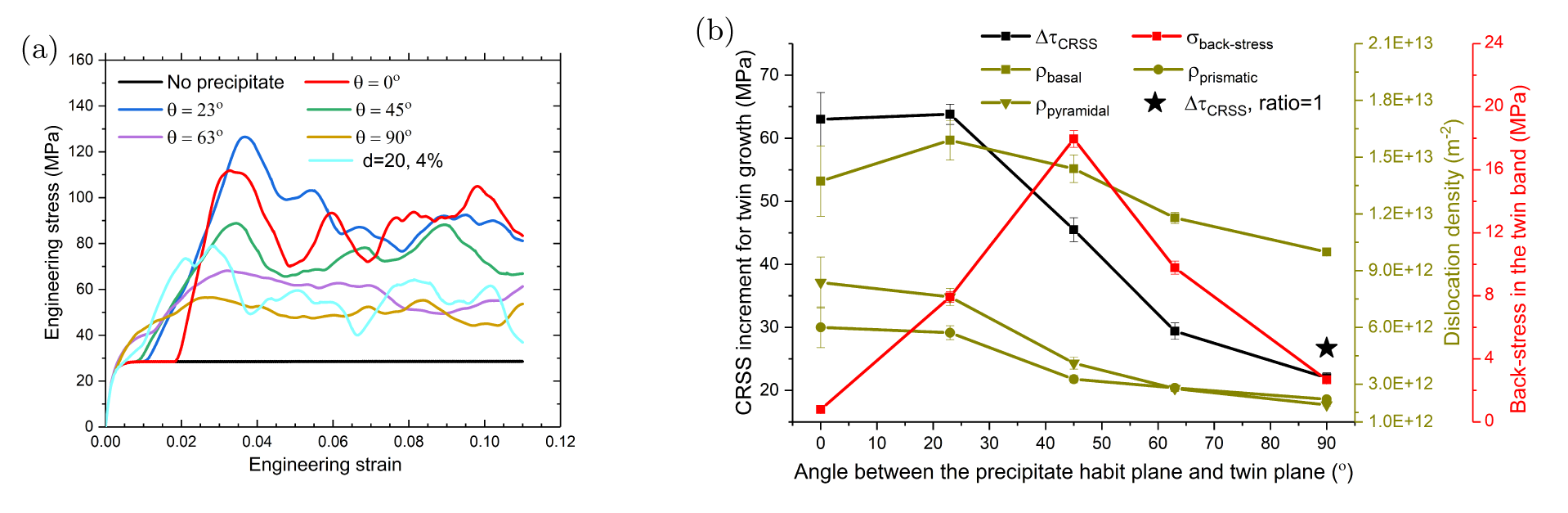}
	\caption{(a) Representative stress--strain curves of the single crystal with various precipitate habit plane. Plate-shaped precipitates with a total volume fraction of \SI{4}{\percent} and an aspect ratio of 10 are randomly distributed in the single crystal. $\theta$ is the alignment angle between the twin plane and the precipitate habit plane. $d$ is the precipitate diameter. 
		(b) Evolution of the CRSS increment ($\Delta \tau_\text{CRSS}$), average basal ($\rho_{\text{basal}}$), prismatic ($\rho_{\text{prismatic}}$), pyramidal ($\rho_{\text{pyramidal}}$) dislocation density in the twinned region, and back-stress ($\sigma_\text{back-stress}$) at a shear strain of \SI{8.3}{\percent} versus the alignment angle between the twin plane and precipitate habit plane. Note that the counterparts in the case containing spherical precipitates with a diameter of 20 elements and the same volume fraction of \SI{4}{\percent} is also given (d=20, \SI{4}{\percent}).
	}
	\label{fig: mechanicalResponses_orien}
\end{figure}

The simulated stress--strain curves of the single crystal containing different oriented plate-shaped precipitates are presented in \cref{fig: mechanicalResponses_orien}a.
All the stress--strain curves in the case with precipitates exhibit similar characteristics, \ie\ multiple stress humps but with different levels of flow stress, indicating the blocking effect of precipitates on twin growth.
The multiple stress drops correspond to the escape of twin boundaries from precipitates. 
Such behavior was also observed in the micro-pillar compression experiments for Mg alloys \citep{Liu2018}. 
Both the mean flow stress and the magnitude of the stress hump decrease with increasing alignment angle between the twin plane and the precipitate habit plane.
The stress--strain curve is much smoother and has the lowest flow stress when the plate habit plane is perpendicular to the twinning plane.

To quantitatively investigate the influence of the habit plane of precipitates on the strengthening effect against twin growth, the mean threshold stress for twin growth in the presence of precipitates is also derived from the present simulations.
It is calculated as the average flow stress between the strain corresponding to the first stress peak and the final strain of \SI{11}{\percent}.
The CRSS increment for twin growth, $\Delta \tau_\text{CRSS}$, is defined as the difference of the threshold stress for twin growth in cases with and without precipitates.  
The predicted $\Delta \tau_\text{CRSS}$ for twin growth in the case of plate-shaped precipitates is plotted in \cref{fig: mechanicalResponses_orien}b as a function of the alignment angle between the twin plane and the precipitate habit plane.
It is obvious that the habit plane of the plate-shaped precipitates has a profound influence on the threshold stress for twin growth assuming the same precipitate volume fraction, number density, and morphology.
The results show that the strengthening effect against twin growth significantly depends on the precipitate habit plane when $\theta > 25^\circ$ while approximately the same strengthening effect for different oriented precipitates was predicted for $\theta < 25^\circ$.
When the angle increases from \SI{22.5}{\degree} to \SI{90}{\degree}, the predicted $\Delta \tau_\text{CRSS}$ substantially decreases from \SI{60}{\mega \pascal} to \SI{22}{\mega \pascal}, which might be attributed to the change of the interparticle spacing on the twin plane, the reduced effective in-contact area and the misfit stress resulted from the strain incompatibility between the sheared matrix and the elastic precipitates.

As the interparticle spacing is inversely proportional to the \term{Orowan} stress required to bow twinning dislocations around the particles, the increase of the interparticle spacing on the twin plane with increasing $\theta$ to some extent leads to the reduction of the threshold stress for twin growth.
However, as the magnitude of the \term{Burgers} vector of the twinning partial dislocation is approximately \num{5} times smaller than that of a basal dislocation, the reduction in the \term{Orowan} stress should only have a minor impact on the decrease of the threshold stress for twin growth.
Instead, the internal misfit stress arising when shear resistant precipitates embedded in the twin is expected to play an important role on the strengthening effect against twin growth.

\cref{fig: orientation}(b-f) show the twinning microstructure, dislocation activities and resolved shear stress on the \hkl{-1012} twin system in the single crystal containing different oriented plate-shaped precipitates at a strain of \SI{8.3}{\percent}.
In all cases, the internal stress distribution is strongly inhomogeneous while the internal stress is substantially higher when the precipitate habit plane is parallel to the twinning plane compared to the case with perpendicular habit plane.
In addition, large detached zones, \ie\ untwinned regions near precipitate surfaces as seen in \cref{fig: orientation}(b$_1$ and b$_2$) were observed when $\theta$ is small.
Profuse prismatic and pyramidal dislocations are activated inside these triangular detached zones, as shown in \cref{fig: orientation}(d and e), which implies that the precipitates with a small alignment angle relative to the twin boundary are highly effective against twin growth. 
The variation of the average basal, prismatic and pyramidal dislocation density in the twin band and also the back-stress at a shear strain of \SI{8.3}{\percent} are plotted in \cref{fig: mechanicalResponses_orien}b as a function of the alignment angle $\theta$.
Here the back-stress is calculated as the difference between the average shear stress in the twin band and the applied shear stress.
It is found that the dependence of dislocation densities (activated to accommodate the strain incompatibility) on the precipitate habit plane is also more profound for $\theta$ ranging from \SI{22.5}{\degree} to \SI{62.5}{\degree}.
However, the back-stress does not follow a similar trend as $\Delta \tau_\text{CRSS}$, it increases with alignment angle  $\theta$ up to \SI{45}{\degree} and then decreases.

Furthermore, the predicted $\Delta \tau_\text{CRSS}$ for twin growth in the case containing spherical precipitates is given in \cref{fig: mechanicalResponses_orien}b.
The same number density and volume fraction as presented above for plate-shaped precipitates was used.
It can be seen that $\Delta \tau_\text{CRSS}$ for twin growth in the case of spherical precipitates is on a similar level compared to that observed for the case of plate-shaped precipitates whose habit plane is perpendicular to the twin plane.
This observation also verifies that the interparticle spacing on the twin plane is a critical material parameter affecting the hardening effect against twin growth.

\subsection{Influence of precipitate volume fraction}
\label{sec: volume_fraction}

\begin{figure}
	\centering
	\includegraphics[width=0.8\textwidth]{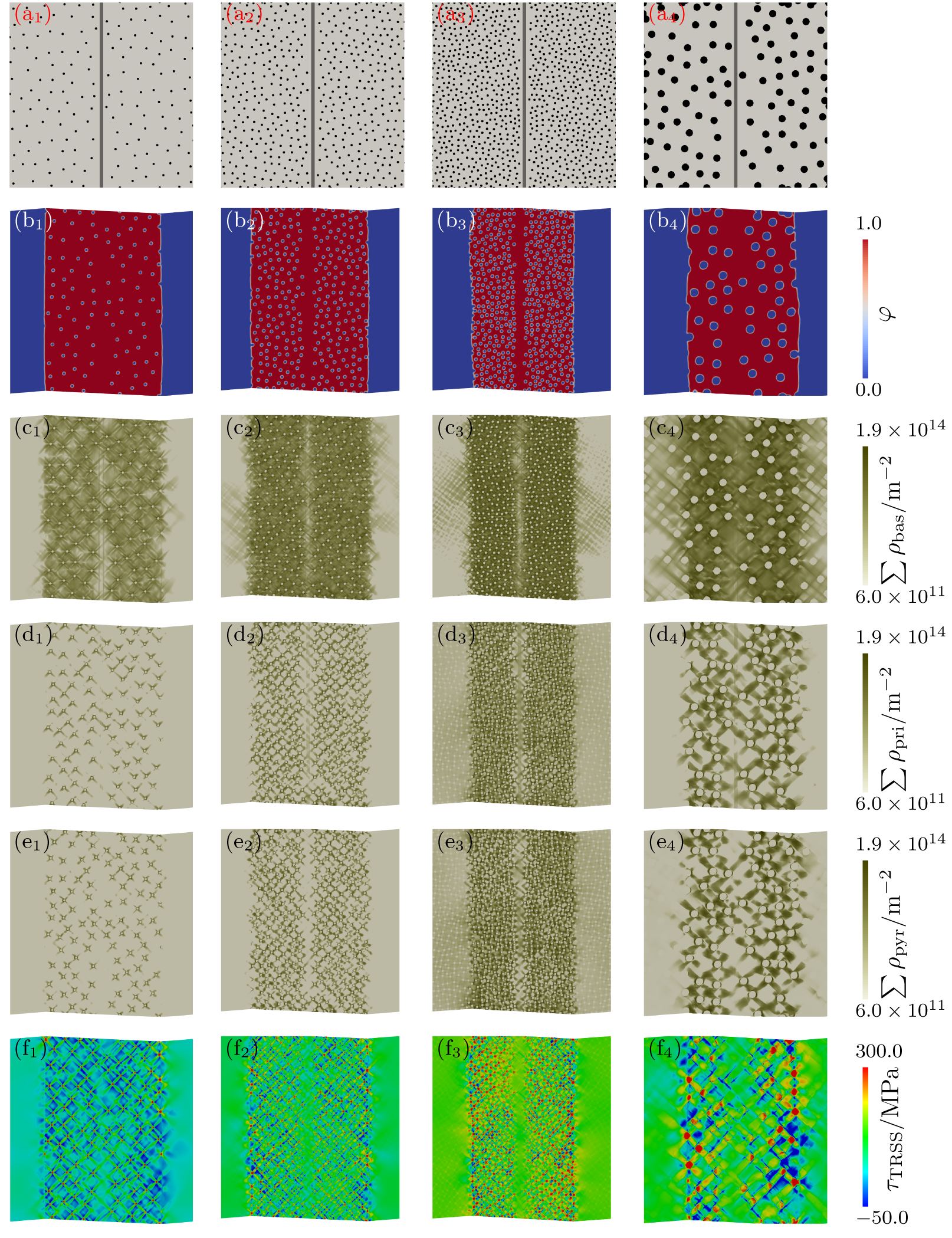}
	\caption{%
		Spatial distribution of spherical precipitates with a uniform diameter of 6 elements and a volume fraction of $(a_1)$ \SI{2}{\percent}, $(a_2)$ \SI{6}{\percent}, and $(a_3)$ \SI{12}{\percent} and $(a_4)$ the counterpart with a diameter of 20 elements and a volume fraction of \SI{12}{\percent}. Precipitates are given in black color and the pre-existing twin band is given in dark grey color.
		The corresponding distribution of the (b) twinning microstructure. The legend symbol $\varphi$ refers to the structure variable, \ie\ $\varphi=0$ within the matrix and $\varphi=1$ within the twin. (c) dislocation density of the basal slip system ($\rho_{\text{bas}}$), (d) dislocation density of the prismatic slip system ($\rho_{\text{pris}}$), (e) dislocation density of the pyramidal slip system ($\rho_{\text{pyr}}$), and (f) resolved shear stress on the twin plane ($\tau_\text{TRSS}$) of the single crystal containing spherical precipitates at a shear strain of \SI{8.3}{\percent}.
	}
	\label{fig: volume_fraction}
\end{figure}

\begin{figure}
	\centering
	\includegraphics[width=1.0\textwidth]{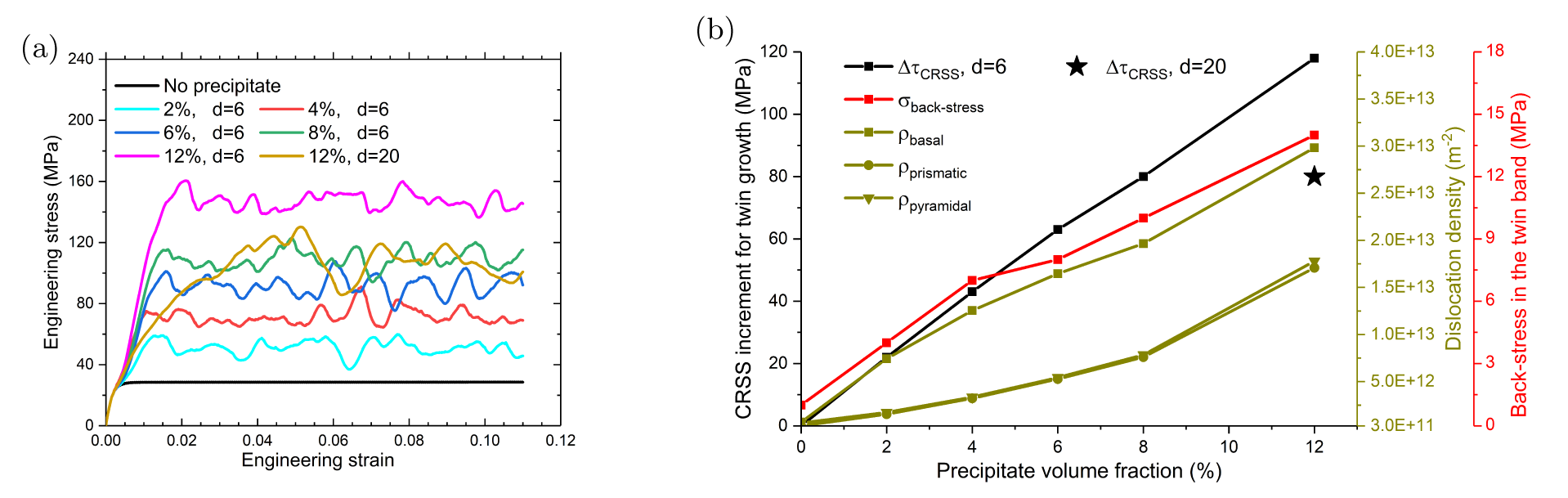}
	\caption{(a) Representative stress--strain curves of the single crystal with various precipitate volume fraction. Spherical precipitates with a uniform diameter of 6 elements are randomly distributed in the single crystal.
		$d$ is the precipitate diameter. 
		(b) Evolution of the CRSS increment ($\Delta \tau_\text{CRSS}$), average basal ($\rho_{\text{basal}}$), prismatic ($\rho_{\text{prismatic}}$), pyramidal ($\rho_{\text{pyramidal}}$) dislocation density in the twinned region, and back-stress ($\sigma_\text{back-stress}$) at a shear strain of \SI{8.3}{\percent} with the precipitate volume fraction. Note that the counterparts with a precipitate diameter of 20 elements and a volume fraction of \SI{12}{\percent} are also given.
	}
	\label{fig: mechanicalResponses_volumeFraction}
	
\end{figure}

The volume fraction of precipitates plays a substantial role on the strengthening effect for both dislocation slip and twin growth and, thus, on the yield stress and mechanical anisotropy of Mg alloys.
In this section, the influence of precipitate volume fraction on the CRSS for twin growth in Mg alloys is quantitatively investigated.
In the simulations, precipitates with a volume fraction of \SI{2}{\percent}, \SI{4}{\percent}, \SI{6}{\percent}, \SI{8}{\percent} and \SI{12}{\percent} are randomly distributed in the single crystal, see \cref{fig: volume_fraction}a.
It is worth to mention that spherical precipitates are selected here to avoid the influence of precipitate morphology.

\cref{fig: mechanicalResponses_volumeFraction}a shows the simulated stress--strain curves of the single crystal with varying volume fractions of precipitates under shear deformation.
The evolution of the CRSS increment of twin growth as a function of precipitate volume fraction is given in \cref{fig: mechanicalResponses_volumeFraction}b.
It can be seen that the predicted CRSS increment increases approximately linearly with the precipitate volume fraction.
The CRSS for twin growth of the single crystal with \SI{8}{\percent} precipitates is \SI{105}{\mega \pascal}, more than \num{3} times higher than that without precipitates (\SI{29}{\mega \pascal}).
The variation of the average basal, prismatic, and pyramidal dislocation densities in the twin band and also the back-stress at a shear strain of \SI{8.3}{\percent} are plotted in \cref{fig: mechanicalResponses_volumeFraction}b as a function of precipitate volume fraction.
As expected, all the dislocation densities increase with increasing precipitate volume fraction. 
However, the increase rate of the dislocation densities varies in different precipitate volume fraction regimes, \eg\ the non-basal dislocation activities increase faster when the volume fraction is above \SI{8}{\percent} as revealed by the increasing slope in \cref{fig: mechanicalResponses_volumeFraction}b.
Although the predicted back-stress increases with increasing precipitate volume fraction, the values of the back-stress are approximately one order of magnitude lower than that of the CRSS increment.
In the current full-field simulations, the back-stress in the case with \SI{8}{\percent} spherical precipitates is simulated to be only $\sim \SI{10}{\mega \pascal}$ considering plastic relaxation, as shown in \cref{fig: mechanicalResponses_volumeFraction}b.
The back-stress estimated by the \term{Eshelby} inclusion model, \ie\ $\sim\gamma fG/2$, is around $\SI{93}{\mega \pascal}$, based on the assumption that accommodation of the strain incompatibility is entirely elastic.
The results show that dislocation activation around precipitates substantially relaxes the back-stress.

It is generally observed that the CRSS increment for twin growth increases with increasing precipitate volume fraction up to the peak aged condition and decreases significantly at the over-aged condition~\citep{Stanford2012,Hidalgo-Manrique2017}, which seems to be in conflict with the predicted linear relationship between the precipitate volume fraction and the strengthening effect against twin growth.
But this can be attributed to the relatively large precipitate size at the over-aged condition compared to that at under-aged and peak-aged conditions.
Another simulation was performed to study the twin growth in the over-aged Mg alloys,  where the diameter of the spherical precipitates was selected to be 20 elements instead of 6 elements with the same volume fraction of \SI{12}{\percent}. 
The corresponding simulated stress--strain curve and the CRSS increment for twin growth are also shown in \cref{fig: mechanicalResponses_volumeFraction}.
The CRSS increment for twin growth substantially decreases from \SI{118}{\mega \pascal} to \SI{80}{\mega \pascal} when the precipitate diameter increases from 6 elements to 20 elements for the same precipitate volume fraction of \SI{12}{\percent}. 
Therefore, the current model agrees well with experimental observations considering the realistic precipitate size and volume fraction subjected to various heat treatments.
\subsection{Influence of precipitate size}
\label{sec: size}

\begin{figure}
	\centering
	\includegraphics[width=0.8\textwidth]{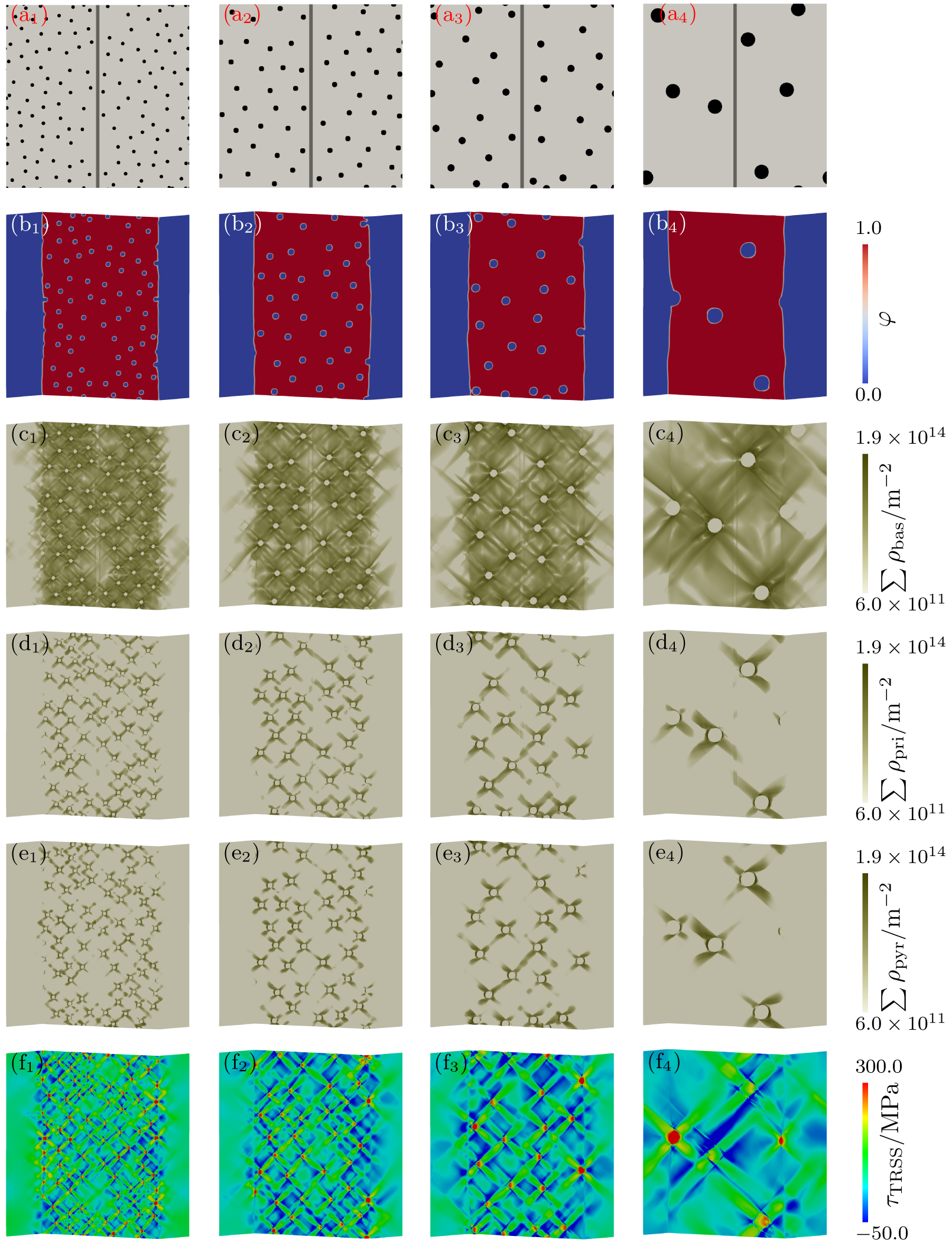}
	\caption{%
		Spatial distribution of spherical precipitates with the same volume fraction of \SI{4}{\percent} but a diameter of $(a_1)$ 10, $(a_2)$ 15, $(a_3)$ 20 and $(a_4)$ 40 elements, respectively.
		Precipitates are given in black color and the pre-existing twin band is given in dark grey color.
		The corresponding distribution of the (b) twinning microstructure. The legend symbol $\varphi$ refers to the structure variable, \ie\ $\varphi=0$ within the matrix and $\varphi=1$ within the twin. (c) dislocation density of the basal slip system ($\rho_{\text{bas}}$), (d) dislocation density of the prismatic slip system ($\rho_{\text{pris}}$), (e) dislocation density of the pyramidal slip system ($\rho_{\text{pyr}}$), and (f) resolved shear stress on the twin plane ($\tau_\text{TRSS}$) of the single crystal containing spherical precipitates at a shear strain of \SI{8.3}{\percent}.
	}
	\label{fig: size}
\end{figure}

\begin{figure}
	\centering
	\includegraphics[width=1.0\textwidth]{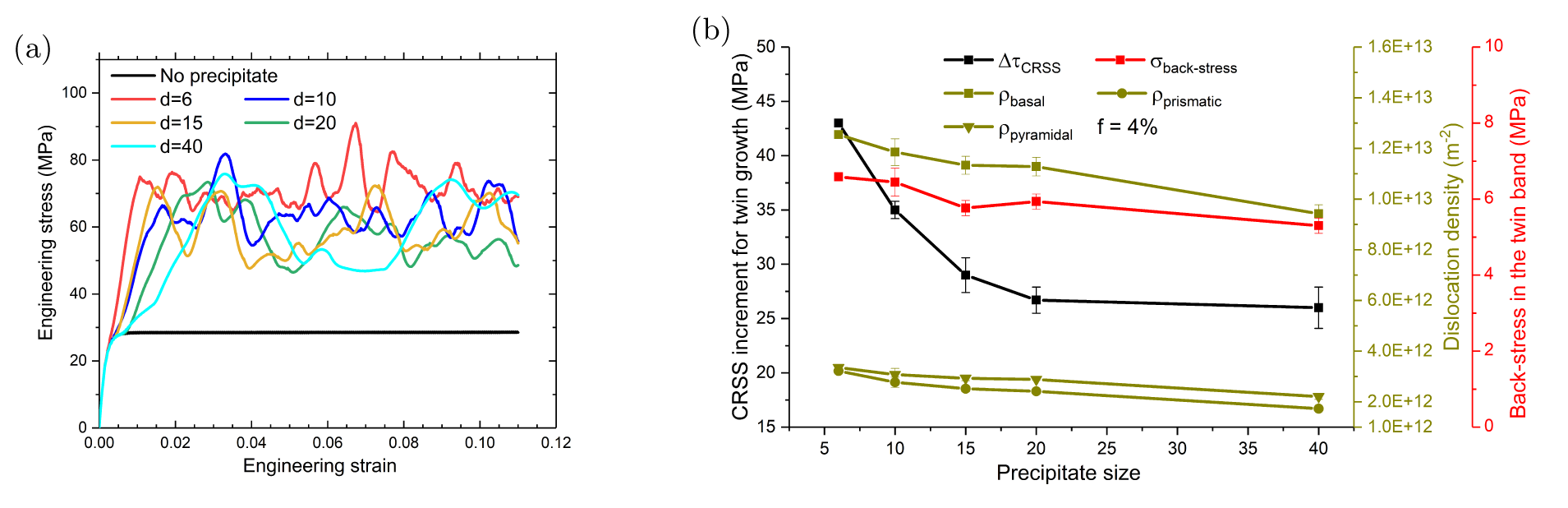}
	\caption{(a) Representative stress--strain curves of the single crystal with various precipitate size and the same volume fraction of \SI{4}{\percent}. $d$ is the precipitate diameter. 
		(b) Evolution of the CRSS increment ($\Delta \tau_\text{CRSS}$), average basal ($\rho_{\text{basal}}$), prismatic ($\rho_{\text{prismatic}}$), pyramidal ($\rho_{\text{pyramidal}}$) dislocation density in the twinned region, and back-stress ($\sigma_\text{back-stress}$) at a shear strain of \SI{8.3}{\percent} as a function of precipitate size.
	}
	\label{fig: mechanicalResponses_size}
\end{figure}

The influence of precipitate size on the threshold of dislocation slip can be quantitatively described by the \term{Orowan} hardening model \citep{Nie2003}.
However, no generally accepted hardening law exists to describe the influence of the precipitate size on the threshold stress for twin growth.
Therefore, this section is dedicated to study the distinct hardening effect against twin growth for precipitates with different sizes at the same volume fraction.
In the simulations, spherical precipitates with a uniform diameter of 6, 10, 15, 20 and 40 elements are randomly distributed in the single crystal at the same total volume fraction of \SI{4}{\percent}, as shown in \cref{fig: size}a.

The simulated mechanical responses of the single crystal with different precipitate sizes are depicted in \cref{fig: mechanicalResponses_size}a.
\Cref{fig: mechanicalResponses_size}b presents the evolution of the CRSS increment for twin growth, dislocation densities, and the mean back-stress in the twin band as a function of the precipitate size at a shear strain of \SI{8.3}{\percent}.
The CRSS increment increases from \SI{28}{\mega \pascal} to \SI{43}{\mega \pascal}, \ie\ it is \SI{54}{\percent} higher when the precipitate size decreases from 20 to 6 elements. 
This reveals a similar strengthening effect of precipitates again twin growth as that against dislocation slip, \ie\ ``smaller is harder''.
\revision{Furthermore, a transition in the strengthening effect of precipitate size on the CRSS for twin growth is observed at a critical precipitate size of 20 elements.
Below this critical size, there is a strong dependence of the strengthening effect on the precipitate size. 
More specifically, a larger number of smaller precipitates gives a higher CRSS increment. 
Above the critical value, there is no size dependence on the CRSS increment.}
It indicates that the precipitate size regime can be split into ``\term{Orowan} stress" regime below the critical precipitate size and ``back-stress" regime above the critical precipitate size.
This observation is consistent with the fact that the distance between precipitates is recognized as the critical material parameter for the \term{Orowan} hardening model while the volume fraction is the key parameter in the back-stress model.
And the back-stress model does not consider the precipitate distribution and interactions among neighboring precipitates.
It is also verified by the simulations that the mean back-stress in the twin band is nearly independent on the precipitate size.
From \cref{fig: mechanicalResponses_size}b it can be seen that it has a constant value of around \SI{6}{\mega \pascal} for a precipitate volume fraction of \SI{4}{\percent}.

\subsection{Influence of precipitate aspect ratio}
\label{sec: shape_ratio}

\begin{figure}
	\centering
	\includegraphics[width=0.8\textwidth]{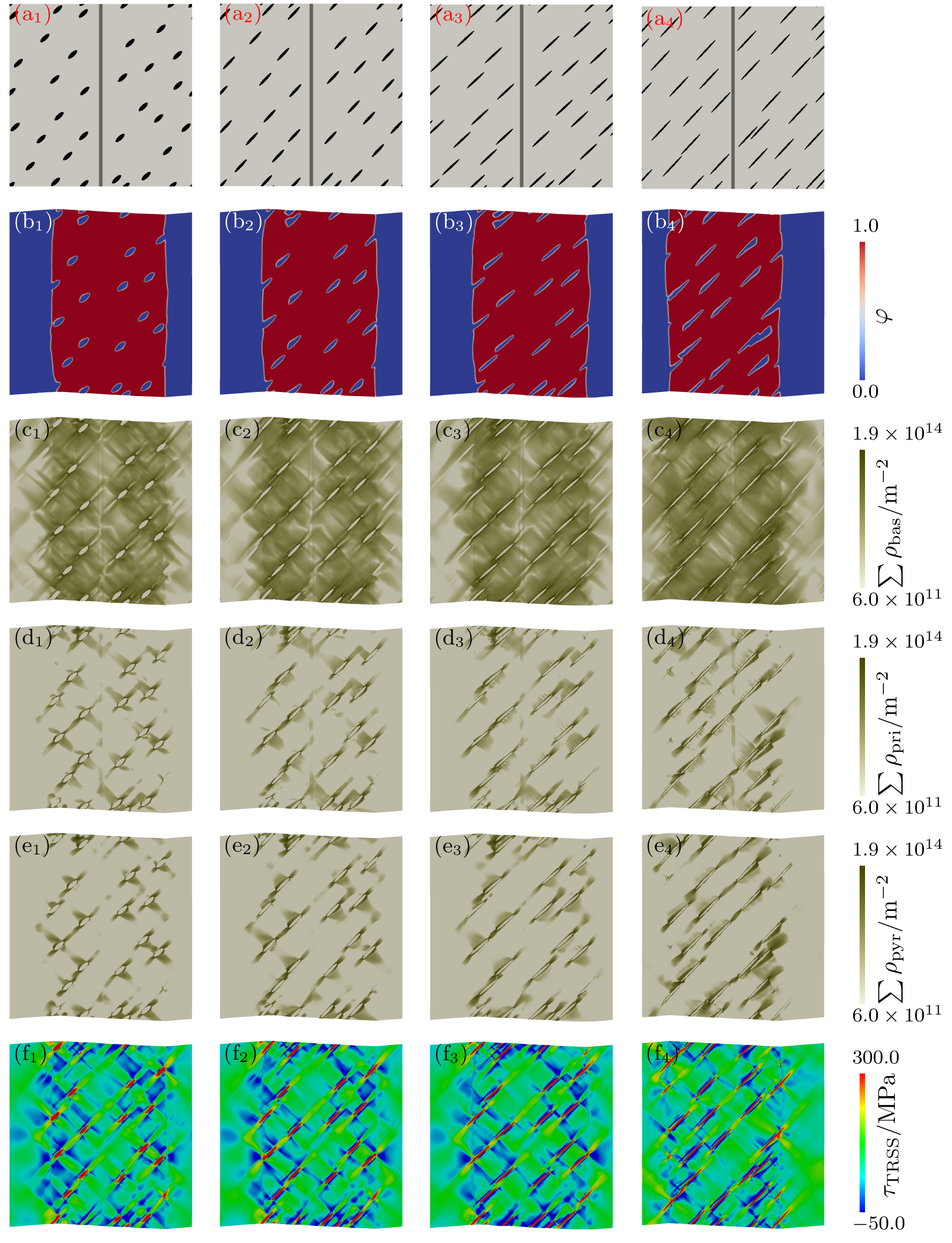}
	\caption{%
		Spatial distribution of basal plate precipitates with the same effective diameter of 20 elements but an aspect ratio of $(a_1)$ 3, $(a_2)$ 6, $(a_3)$ 10 and $(a_4)$ 16, respectively. 
		The total volume fraction of the basal plate precipitates is \SI{4}{\percent}.
		Precipitates are given in black color and the pre-existing twin band is given in dark grey color.
		The corresponding distribution of the (b) twinning microstructure. The legend symbol $\varphi$ refers to the structure variable, \ie\ $\varphi=0$ within the matrix and $\varphi=1$ within the twin. (c) dislocation density of the basal slip system ($\rho_{\text{bas}}$), (d) dislocation density of the prismatic slip system ($\rho_{\text{pris}}$), (e) dislocation density of the pyramidal slip system ($\rho_{\text{pyr}}$), and (f) resolved shear stress on the twin plane ($\tau_\text{TRSS}$) of the single crystal containing basal plate precipitates at a shear strain of \SI{8.3}{\percent}.
	}
	\label{fig: ratio}
\end{figure}

\begin{figure}
	\centering
	\includegraphics[width=1.0\textwidth]{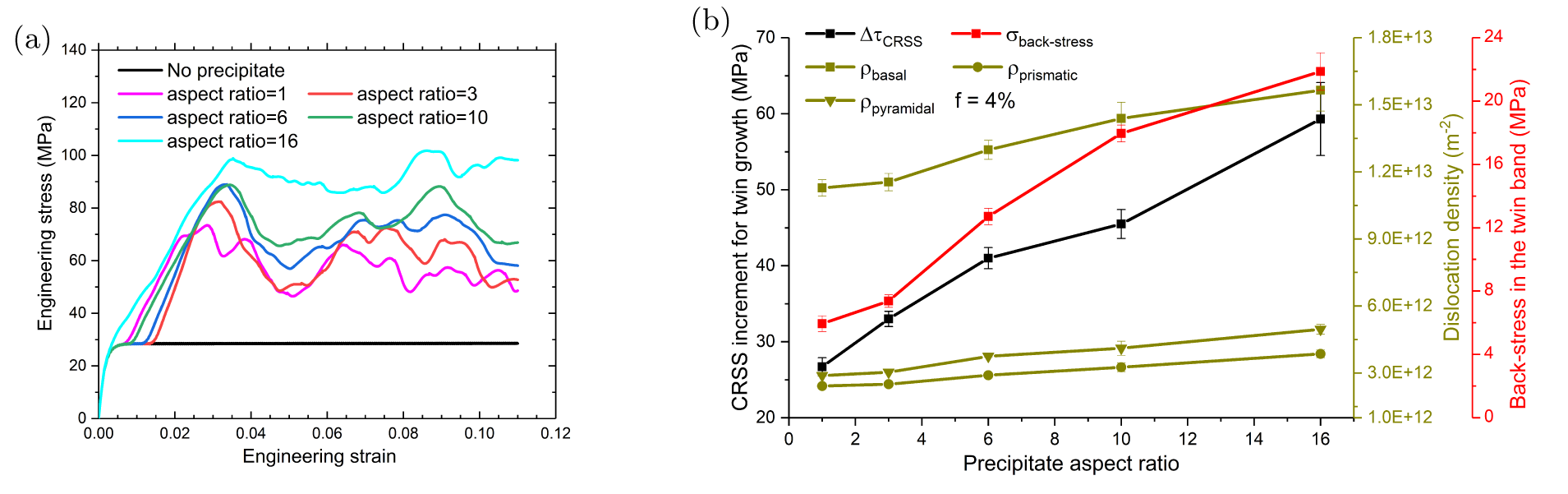}
	\caption{(a) The influence of the precipitate aspect ratio on the mechanical responses of the single crystal with \SI{4}{\percent} basal plate precipitates. The effective diameter of the plate precipitates is 20 elements. 
		(b) Evolution of the CRSS increment ($\Delta \tau_\text{CRSS}$), average basal ($\rho_{\text{basal}}$), prismatic ($\rho_{\text{prismatic}}$), pyramidal ($\rho_{\text{pyramidal}}$) dislocation density in the twinned region, and back-stress ($\sigma_\text{back-stress}$) at a shear strain of \SI{8.3}{\percent} versus precipitate aspect ratio.
	}
	\label{fig: mechanicalResponses_ratio}
\end{figure}

The high strength of precipitation hardened aluminum alloys---as an example for a well-studied reference material class---is typically ascribed to both the high number density and the large aspect ratio (usually above 40:1) of $\hkl{111}_\alpha$ plate-shaped precipitates \citep{Ardell1985,Bardel2014}. 
In Mg alloy design, therefore much effort has also been placed on achieving a high precipitate number density by adjusting composition and optimizing the associated heat treatments \citep{Nie2012}.
However, the exploration on the influence of the precipitate aspect ratio on the hardening effect against twinning in Mg alloys still remains limited.
This section aims at quantifying the influence of the precipitate aspect ratio on the CRSS of twin growth in Mg alloys. 
Assuming the same volume for each single precipitate (equivalent diameter = 20 elements) and the identical precipitate number density, basal plate precipitates with different aspect ratios are randomly distributed in the single crystal. 
The plate aspect ratio varies from 1 to 16 and the total precipitate volume fraction is \SI{4}{\percent}, as shown in \cref{fig: ratio}a.

\cref{fig: mechanicalResponses_ratio}a shows the simulated stress--strain curves of the single crystal with different plate aspect ratios under shear deformation.
The variations of the CRSS increment, dislocation densities and the average back-stress in the twin band as a function of the precipitate aspect ratio are shown in \cref{fig: mechanicalResponses_ratio}b at a shear strain of \SI{8.3}{\percent}.
\cref{fig: mechanicalResponses_ratio}a shows that the stress--strain curves are shifted upwards as the plate aspect ratio deviates from 1.
Thus, the plated-shaped precipitates are more effective at blocking twin growth, as compared to spherical precipitates of the same size.
\cref{fig: mechanicalResponses_ratio}b presents that the CRSS increment for twin growth increases almost linearly with increasing the plate aspect ratio. 
The CRSS increment increases from \SI{28}{\mega \pascal} to \SI{55}{\mega \pascal} (almost two times higher) when increasing the plate aspect ratio from 1 to 16 at the same precipitate size and volume fraction.
A linear dependence of the mean back-stress on the precipitate aspect ratio is also found.
The mean back-stress induced in the single crystal with spherical precipitates is only \SI{4}{\mega \pascal}, whereas it is \SI{20}{\mega \pascal} when the precipitate aspect ratio reaches 16.
It can be seen that the average dislocation densities also increase with increasing the precipitate aspect ratio, which means that the strain incompatibility for plate-shaped precipitates during twin-precipitate interaction process is larger than in the case of spherical precipitates. 
These dislocations are required to relax the misfit stress that develops as a result of the strain incompatibility between the sheared matrix and elastically deformed precipitates.

\section{Discussion}
\label{sec: discussion}

This work has employed an integrated crystal plasticity-phase field framework to predict and help understanding the influence of precipitates on growth of a pre-existing $\{10\bar{1}2\}$ twin in Mg alloys. 
The model has allowed the details of the twin-precipitate interaction to be studied and also provides estimates of the increment in CRSS for twin growth due to precipitates. 
Growth (thickening) is the final step in the twinning process and for comparing these results obtained here with experiments it is first necessary to briefly consider the complete twinning process and the influence precipitates can have on the preceding steps. 
For a more detailed discussion, the reader is referred to a recent review on twin-particle interactions in Mg alloys \citep{Robson2018}.

Twinning can be considered in three stages: nucleation, propagation (of the twin front across a grain), and growth (thickening of the twin).
Nucleation occurs by the simultaneous formation of multiple twinning partial dislocations and accompanying atomic shuffles in a region of stress concentration, most commonly at grain boundaries \citep{WangJ_etal2009,Wang_etal2010_3}.
Experimental evidence suggests that the stress required to nucleate a twin is not strongly affected by the presence of precipitates \citep{Robson2018}.
Propagation involves the migration of the twin front across the grain until it is arrested or impinges on the opposite grain boundary. 
This can be considered as the movement of a super-dislocation wall producing the twinning shear. 
In Mg, most twins are observed to completely span grains, even when precipitates are present. 
Therefore, it is usually assumed that the stress required for propagation is much lower than that required for nucleation and is easily exceeded by the mean stress in the grain. Based on this assumption, the role of precipitates on the propagation stage is usually ignored. 
Recently, however, Barnett et al. \citep{Barnett2019,Barnett2019_2} have shown that this may not be a valid assumption and propagation may also play a critical role in controlling the stress at which twinning impacts on the material response. 
The final stage of twinning involves growth (\ie\ thickening) of the twin. 
In this case, additional twinning dislocation loops nucleate and expand leading to twin growth perpendicular to propagation direction. 
The increase in required stress for this growth stage is usually considered as most important in estimating the effect of precipitates on the stress for twinning \citep{Robson2018}. 
As demonstrated by Barnett \citep{Barnett2019}, increasing the CRSS for propagation or the CRSS for growth will lead to an increase in the twinning stress, but with a slightly different functional relationship.

Experimentally derived twinning stress values for polycrystalline Mg are typically determined by measuring the proof stress (\eg\ at 0.1\% offset strain) and then using an effective, \ie\ average \term{Taylor} factor to account for the influence of texture. 
Alternatively, a crystal plasticity model can be used to adjust the CRSS values for each deformation mode until best fit is achieved between the predicted and measured flow curves. 
Both of these approaches are subject to significant error and furthermore the effective CRSS for twinning derived in this way will correspond to the stress required to achieve clearly measurable plastic strain, which necessarily requires not only twin nucleation and propagation but also growth \citep{Stanford2009}. 
Therefore, the experimental measurements are usually influenced by all three stages of twinning, with twin growth dominant beyond the initial yield point. 

For plate-shaped precipitates on the basal plane experiments on Mg-Al alloys have reported increases in the CRSS for twinning in the range 5--50$\,$MPa (\citep{Robson2018},Table 1). 
The lowest of these values was for material in a very overaged condition where the precipitates were large compared to the twin width. 
The highest value was for a structure containing discontinuous precipitates. 
Since neither of these apply in the present case, it is useful to consider only data from the peak aged condition where values in the range 31-46$\,$MPa are reported \citep{Robson2018}. 
Note that even in peak-aged condition, precipitates in Mg-Al alloys are quite large and widely spaced \citep{Nie2012} and the classically calculated \term{Orowan} bowing stress for a twinning dislocation is small (1-7$\,$MPa \citep{Robson2018}).  
As demonstrated in Figure \ref{fig: mechanicalResponses_singlePre}, the model predicts an additional stress of 38$\,$MPa for basal plate precipitates, which falls within the range of the experimental measurements. 

Qualitatively, the predictions also show features that match well with experimental observation. 
The twin boundary is seen to deflect as it becomes pinned at the particle (\eg\ \cref{fig: evolutionSiglePre}(a$_2$)). 
Eventually, the twin engulfs the particle, and this leads to plastic relaxation by local slip to accommodate the strain incompatibility between the sheared matrix (twin) and unsheared particle \citep{Robson2016}. 
This occurs not only by basal slip, but also prismatic and pyramidal slip. 
The most intense slip activity has been predicted to occur in the matrix ahead of the precipitate tips, where the strain incompatibility is highest, and this is consistent with the observations of Gharghouri et al. \citep{Gharghouri1998}. 
\revision{Moreover, a narrow untwinned region is observed on the downstream side of the precipitate, which is formed by the constraint imposed by the non-shearing precipitate.
Similar observations have been made in experiments \citep{Stanford2012} and MD simulations \citep{Fan2018}.
The area of the untwinned region reflects the strengthening effect of the precipitates on the twin growth, with larger untwinned regions and non-basal dislocation activity observed when the habit plane of precipitates is parallel to the twinning plane (as shown in \cref{fig: orientation}).}

The model predicts that the maximum inhibition occurs at the point the twin front is close to detaching from the particle (\ie\ just before the particle is fully engulfed). 
This is consistent with the idea that at this position the strain incompatibility is at a maximum, since once the precipitate is fully engulfed it undergoes a rigid body rotation \citep{Gharghouri1998}. 
Finally, it is predicted that even after the twin has fully engulfed the particle there will persist a CRSS increase for continued growth. 
This stress increment can be attributed to the inhomogeneous distribution of the stress field and induced dislocation hardening when a precipitate is present. 
Importantly, it is predicted that this stress ($\sim$ \SI{6}{\mega \pascal}) is only around 15\% of the maximum CRSS increase (at the point of particle engulfment as described above). 
In the framework of a classical strengthening model, this demonstrates that most of the CRSS increase is attributed to the local interaction at the point the twin front overcomes the precipitate and not the long range back-stress produced by the incorporation of an unsheared particle inside sheared (twinned) material.

The good agreement between the model and experimental observations, both quantitatively in prediction of the CRSS increment and qualitatively in reproducing the observed plastic relaxation, 
gives confidence in making predictions for the important parameters that can be manipulated by alloy design: precipitate orientation, precipitate volume fraction, precipitate size, and precipitate shape (aspect ratio). 

\begin{itemize}
\item \textbf{Effect of precipitate habit plane:}\
Precipitate (plate) orientation has previously been predicted to have a strong effect on strengthening against twin growth, whether calculated on the basis of classical models (\term{Orowan} or back-stress based \citep{Robson2018}) or MD simulations \citep{Fan2018,Fan2018b}. 
Consistent with previous MD simulations, the present work demonstrates that the maximum strengthening occurs when the plate habit plane is parallel to the twinning plane. 
The minimum strengthening effect occurs when the habit plane is perpendicular to the twinning plane. 
The present work and previous MD simulations predict a difference of about 40$\,$MPa between the ``hard'' and ``soft'' precipitate orientations \citep{Fan2018b}. 
Plates with a basal or prismatic habit plane (the most common in practice) are predicted to provide similar levels of strengthening, close to the median of the minimum and maximum. 
Orienting the habit plane parallel to the twinning plane (the ``hard'' orientation) maximizes the area of twin boundary that has to overcome the precipitate as the twin engulfs the particle. 
This leads to the maximum inhibition of twin growth. 
The simulation results also revealed that in this ``hard'' orientation, extensive basal and non-basal slip is triggered. 
This is a further consequence of the high stress level needed to sustain twin growth. 
It is also an important effect which might help to achieve a more homogeneous distribution of slip activity, more activation of secondary slip, less stress concentration and thus higher overall ductility.

\item \textbf{Effect of precipitate volume fraction:}\
Increasing precipitate volume fraction at a constant precipitate size is predicted to lead to a strong increase in CRSS for twin growth. 
The relationship is almost proportional up to a volume fraction of 12\% (the maximum investigated in this study). 
Increasing volume fraction is predicted not only to increase the unrelaxed back-stress (as would be predicted from a classical \term{Eshelby} calculation) but also increases the dislocation activity (plastic relaxation) required to sustain twin growth. 
However, the simulations also reveal that precipitate size and spacing can be important parameters in controlling the CRSS increment. 
Comparing the same volume fraction, larger and more widely spaced precipitates give lower strengthening. 
This is consistent with experimental observations of a reduction in strengthening against twin growth in the overaged state \citep{Robson2018}.
 Note that in the present simulations, even in the overaged case, the precipitates are much smaller than the twin width and the twin is forced to engulf them to continue growth. 
 In very overaged Mg-Al alloys, the precipitates can become very large and widely spaced with respect to the twin width. 
 In such cases, twins are observed to navigate through a precipitate containing microstructure through deflection of the boundary plane to avoid engulfing the precipitates. 
 Therefore, the spatial distribution of the precipitates becomes very important, since this determines the size and arrangement of the precipitate free channels.

\item \textbf{Effect of precipitate size:}\
A detailed investigation of the effect of precipitate size (at a fixed volume fraction) on CRSS for twin growth has revealed that there is a transition in behaviour at a critical precipitate size (Figure~\ref{fig: mechanicalResponses_size}). 
Below this critical size, there is a strong dependence of the strengthening effect on the precipitate size.
More specifically, a larger number of smaller precipitates gives a higher CRSS increment (\term{Orowan}-like behavior). 
Above the critical value, there is no size dependence on the CRSS increment (\term{Eshelby}-like behavior). 
The prediction that both types of behavior are possible depending on particle size solves a previous contradiction in the literature between studies proposing an \term{Orowan}-like strengthening effect as most efficient and other studies proposing a back-stress (\term{Eshelby}-like) strengthening response \citep{Robson2018}.

\revision{It is worth noting that very small and coherent precipitates can be sheared by tensile twins in Mg alloys, thus suggesting an upper limit to the Orowan-like hardening behavior.
It has been reported in a Mg-6Gd-1Zn-0.6Zr (wt\%) alloy that the thin and coherent $\gamma'$ basal plate precipitates lead to a strengthening of only around 10 MPa \citep{Geng2011}.
TEM analysis revealed that these particles were sheared by tensile twins, which suggests that they provide little resistance to the twin growth.
However, the common precipitate types in Mg-RE alloys are plate precipitates on the prismatic plane, \textit{e.g.} $\beta_1-\text{Mg}_3(\text{Gd,Y})$ precipitates in Mg-Gd-Y alloys \citep{Nie2012}. 
These larger prismatic plate precipitates, obtained in the peak-aged state, are resistant to shearing by tensile twins \citep{Liu2018}.
Moreover, recent experimental and MD studies have indicated that precipitates are usually not sheared in commercially important Mg alloys, such as Mg-Al \citep{Stanford2012,Fan2018} and Mg-Zn \citep{Stanford2009,Jain2015} alloy series.
In the current work, the precipitates are assumed to be non-shearable, which correspond to the most common case of twin-precipitate strengthening.
However, it is expected that this assumption will break down in the case of very small precipitates, where the effect of precipitate shearing on strengthening remains to be studied.}

\item \textbf{Effect of precipitate aspect ratio:}\
The final parameter to be studied using the model was the precipitate aspect ratio. 
The present results predict that the CRSS increment should increase with an increase in precipitate aspect ratio for the basal plates studied here. 
As for the effect of volume fraction, this increase is predicted to be accompanied by an increase in the unrelaxed back-stress and an increase in dislocation activity within the twin. 
The effect of aspect ratio was also explored in previous investigations where an analytical model to predict the CRSS increment for twin growth due to precipitates was developed \citep{Fan2018}. 
This model also predicts an increasing aspect ratio on CRSS increment over most of the range studied in this work. 
The functional form of the relationship predicted by the two models is somewhat different, with the previous work suggesting a maximum in the strengthening at an aspect ratio of around 10, whereas our current simulations suggest an increasing strengthening effect up to at least an aspect ratio of 16. 
The reason for this discrepancy is probably due to a difference in the way the plastic relaxation of the back-stress is handled. 
In the analytical model, a constant (50\%) relaxation of the elastic back-stress was assumed. 
The present model, in which the plastic relaxation process is directly simulated rather than approximated, predicts that the amount of plastic relaxation is itself dependent on the precipitate aspect ratio, which was not accounted for in previous studies.
However, there is not yet systematic experimental evidence to assess the effect of aspect ratio on the maximum strengthening effect against twin growth.
Such experiments are challenging since the aspect ratio of precipitates cannot be readily manipulated.
\end{itemize}

\section{\revision{Conclusions and outlook}}
\label{sec: conclusion}

In this work, a full-field coupled CP-PF model has been employed to study the interaction between twin growth and precipitates in Mg alloys.
The twinned phase is represented by a non-conserved order parameter, and its evolution, \ie\ twin growth, is described by a \term{Ginzburg}--\term{Landau} equation, whereas the concurrent plastic deformation occurring in  twinned and un-twinned phases are described by a dislocation-density-based CP model. 
Full-field simulations were performed to systematically investigate the influence of precipitate orientation, size, volume fraction, and shape (aspect ratio) on the interaction between precipitates and a growing twin, including plastic relaxation processes. 
All simulations reveal that local plastic relaxation by basal, prismatic and pyramidal slip occurs as a precipitate enters a twin. 
The maximum resistance to twin growth occurs at the point the precipitate is partly embedded in the twin prior to undergoing rigid body rotation. 
Even once the precipitate is fully embedded in the twin and no longer in contact with the twin boundary, increase in CRSS is required for continued growth due to the plastic relaxation and redistribution of stresses.

In addition to understanding the interactions that occur when a twin boundary grows through a precipitate containing microstructure, the simulations were also used to quantitatively predict the increment in the effective CRSS for twin growth due to non-shearing precipitates.

The following conclusions are drawn:
\begin{itemize}

\item The CRSS increments predicted by the model agree well with those determined from experiments and fall within the range of reported measurements.

\item For plate-shaped precipitates, orientation with respect to the twin boundary is a critical parameter in determining the strengthening effect. Precipitates with their habit plane parallel to the twinning plane are predicted to provide the maximum strengthening effect and those perpendicular the minimum effect. Plates with basal or prismatic habit planes provide a similar (median) level of strengthening. There is approximately a doubling of strengthening between the orientations that give the minimum and maximum effects.

\item Increasing precipitate volume fraction at a constant precipitate size is predicted to produce a proportional increase in CRSS increment for twin growth and is thus an effective strategy for strengthening.

\item For a fixed volume fraction, the effect of precipitate size falls into two regimes. Below a critical size, \term{Orowan}-like behaviour is predicted where a reduction in size (and spacing) produces a strong increase in CRSS increment. Above this size, the CRSS increment is size independent (\term{Eshelby}-like behaviour). This explains the previous apparent contradiction in the literature regarding the most appropriate strengthening law to rationalize experimental observations.

\item Increasing the aspect ratio of basal plate-shaped precipitates is predicted to increase the CRSS increment for twin growth. The strong effect of aspect ratio on plastic relaxation behaviour and the level of unrelaxed back-stress is important in explaining this behavior.
\end{itemize}

The model presented here is a useful tool in guiding alloying development to resist twin growth. For maximum strengthening, the model predicts a high volume fraction of small, high aspect ratio, shear-resistant plate-shaped precipitates are desirable. 
Precipitates on either the basal or prismatic planes are predicted to be similarly effective. 
Such precipitates will lead to intense dislocation activity inside the twin due to plastic relaxation processes, which will in turn have a significant effect on any further deformation of the twinned region in polycrystalline alloys under complex boundary conditions. 

\revision{Although the current study has focused on the mechanisms of twin-precipitate interaction in Mg alloys, the model is also applicable to the other HCP materials.
As discussed in the present work, the twin-precipitate interaction mechanisms also depend on the crystallographic slip of the specific material, in addition to the morphologies and distributions of precipitates.
With further comparative studies on the twin-precipitate interactions between other HCP metals, the influence of the anisotropic elastic stiffness and crystallographic slip on the twin-precipitate interaction mechanisms in Mg, Ti, and Zr alloys can be revealed.}

\section{Acknowledgements}
This work was supported by National Key Research and Development Program of China (Grant No. 2016YFB0301103), the National Nature Science Foundation of China (Grant No. 51701117) and the China Postdoctoral Science Foundation (Grant No. 2017M611558).
P.~Shanthraj and J. D.~Robson thanks the EPSRC for financial support through the associated programme grant LightFORM (EP/R001715/1). 
P.~Shanthraj is also grateful to the Airbus\textendash University of Manchester Centre for Metallurgical Excellence for supporting aspects of this research.
C.~Liu acknowledges the kind supports from the State Scholarship Fund of Chinese Scholarship Council (CSC, No.~201506230039).
The data presented in this paper may be obtained by contacting the corresponding author.

\section{References}
\bibliographystyle{elsarticle-num}

\appendix 
\section{Plastic flow relations}
\label{App: Plastic flow relations}
The shear rate of mobile dislocations on each of the $N_\text{s}$ slip systems is described by the \term{Orowan} equation \footnote{Here the \term{Orowan} equation relates the plastic shear strain rate with dislocation density and velocity and is different with the \term{Orowan} hardening model in the introduction, which describes the precipitate hardening against dislocation gliding.} \citep{Orowan1934}:
\begin{equation}\label{eq:Orowan}
\dot{\gamma}  = \rho_\text{m} b v,
\end{equation}
where $\rho_\text{m}$ is the density of mobile dislocations, $b$ the magnitude of the \term{Burgers} vector and $v$ the mobile dislocation velocity. 

Following the thermally activated model firstly proposed by Kocks et al. \citep{Kocks_etal1975}, the dislocation velocity is expressed as
\begin{equation}\label{eq:Rhov}
v=v_0 \operatorname{exp} \left[ - \frac{Q_\text{a}}{\kB T} \left\{1 - \left( \frac{|\tau_\text{eff}|}{\tau_\text{P}} \right)^p \right\}^q\right]\text{sign}(\tau),
\end{equation}
where $v_0$ is the reference dislocation glide velocity, $Q_\text{a}$ is the activation energy required to overcome the obstacles for dislocation motion, $\kB$ is the \term{Boltzmann} constant, $T$ is the absolute temperature, $p$ and $q$ are the parameters describing the impact of applied stress on the activation energy, $\tau_\text{eff}$ is the effective resolved shear stress acting as the driving force for dislocation motion, and $\tau_\text{P}$ is the \term{Peierls} stress.

The effective shear stress $\tau_\text{eff}$ is defined as the resolved shear stress reduced by the passing stress $\tau_\text{pass}$ :
\begin{equation}\label{eq:Taueff}
\tau_\text{eff} = \left\{ \begin{array}{lcl}
|\tau| - \tau_\text{pass} & \mbox{for} & |\tau| > \tau_\text{pass} \\ 
0 & \mbox{for} & |\tau| \leq \tau_\text{pass} 
\end{array}\right.
\end{equation}
The passing stress on the system $\alpha$ is given by :
\begin{equation}\label{eq:Taupass}
\tau^\alpha_\text{pass} = G b \left( \sum_{\alpha'=1}^{N_\text{s}} \varsigma_{\alpha \alpha'} \left( \rho^{\alpha'}_\text{m} + \rho^{\alpha'}_\text{d}\right)\right)^{1/2},
\end{equation}
with the shear modulus $G$, $\varsigma_{\alpha \alpha'}$ describing the interaction strength between dislocations on slip systems $\alpha$ and $\alpha'$, and the dipole dislocation density $\rho_\text{d}$ on each slip system.

The evolution of the mobile dislocation density is determined by dislocation multiplication and annihilation:
\begin{equation}\label{eq:RhoeRate}
\dot{\rho}_\text{m} = \frac{|\dot{\gamma}|}{b \Lambda} - \frac{2\widehat{d}}{b} \rho_\text{m}|\dot{\gamma}| - \frac{2\check{d}}{b} \rho_\text{m}|\dot{\gamma}|.
\end{equation} 
The first term on the right hand describes the increase of the mobile dislocation density due to immobilisation, the second and third term represent the dislocation density decrease caused by dipole formation and dislocations annihilation, respectively.  
$\Lambda$ is the mean free path of the mobile dislocations, $\widehat{d}$ the maximum distance between two glide planes of dislocations that allow formation of a stable dipole and $\check{d}$ the minimum glide plane separation below which the dislocations spontaneously annihilate.

The full set of used material parameters is given in \cref{Tab: matepara,Tab: interaction} \citep{Liu_etal2018}. 

\begin{table}[h!]
	\centering
	\caption{Input material parameters for the constitutive models}
	\resizebox{0.9\linewidth}{!}{%
		\begin{tabular}{ccccccccccc}
			\toprule
			\textbf{Crystal plasticity model} & $b$ (m)& $v_0$ (m/s) & $Q_a $ (J) & $\tau_P$ (Pa) &$d_g$ (m)&$C_{anni} $& p & q & c\\
			\midrule		
			Basal slip systems     & $3.20 \times 10^{-10}$ & $1.0\times10^{-5}$ & $7.0\times10^{-20}$ & $1.5\times10^{7}$  & $2.0\times10^{-5}$ &11.0 & 1.0 & 1.0 &12.0\\
			Prismatic slip systems & $3.20 \times 10^{-10}$ & $1.0\times10^{-5}$ & $7.0\times10^{-20}$ & $7.3\times10^{7}$  & $2.0\times10^{-5}$ &11.0 & 1.0 & 1.0 &12.0\\
			Pyramidal slip systems & $6.11 \times 10^{-10}$ & $1.0\times10^{-5}$ & $7.0\times10^{-20}$ & $1.15\times10^{8}$ & $2.0\times10^{-5}$ &11.0 & 1.0 & 1.0 &12.0\\
			\midrule
			\textbf{Twin phase field model} &$\kappa_{11}/(\Delta f l_0^2)$ & $\kappa_{22}/(\Delta f l_0^2)$ & $\kappa_{33}/(\Delta f l_0^2)$ & $M/ \Delta f $ $(s^{-1})$ 
			& $\Delta f$                & $l_0$\\
			\midrule
			&1.0                           & 1.0                           & 1.0                            & 5.0            
			& $4.4\times10^{8}$            & 1.0\\
			\midrule
			\textbf{Elastic constants}& $C_{11}(Pa)$        & $C_{33}(Pa)$        & $C_{44}(Pa)$        & $C_{12}(Pa)$        & $C_{13}(Pa)$\\
			\midrule
			& $5.93\times10^{10}$ & $6.15\times10^{10}$ & $1.64\times10^{10}$ & $2.57\times10^{10}$ & $2.14\times10^{10}$ \\     
			\bottomrule
	\end{tabular}}
	\label{Tab: matepara}
\end{table}

\begin{table}[h!]
	\centering
	\caption{Interaction strength of dislocations on the slip system $\alpha$ and $\alpha'$}
	\begin{tabular}{cccc}
		\toprule
		$\varsigma_{\alpha \alpha'}$& Basal slip systems & Prismatic slip systems & Pyramidal slip systems\tabularnewline
		\midrule
		Basal slip systems     & 0.8 & 0.8 & 0.8\tabularnewline
		Prismatic slip systems & 0.8 & 1.2 & 1.8\tabularnewline
		Prismatic slip systems & 1.8 & 1.8 & 1.8\tabularnewline
		\bottomrule
	\end{tabular}
	\label{Tab: interaction}
\end{table}
\section{Notation}
As a general scheme of notation, vectors are written as boldface lowercase letters (\eg\ \vctr{a}, \vctr{b}), second-order tensors as boldface capital letters (\eg\ \tnsr{A}, \tnsr{B}), and fourth-order tensors as blackboard-bold capital letters (\eg\ \tnsrfour{A}, \tnsrfour{B}). 
For vectors and tensors, Cartesian components are denoted as, respectively, $a_i, A_{ij} \text{ and } A_{ijkl}$. 
Second-order tensors are represented in this work as linear mappings between vectors and  are denoted as \tnsr{A}\vctr{b} (in components $A_{ij}b_j$, implicit summation over repeated indices is used unless specified otherwise) and, likewise, fourth-order tensors represent linear mappings between second-orders and are designated as \tnsrfour{A}\tnsr{B} ($A_{ijkl}B_{kl}$). 
The composition of two second-order tensors is denoted as \tnsr{A}\tnsr{B} ($A_{ik}B_{kj}$). 
The tensor (or dyadic) product between two vectors is denoted as $\vctr{a}\otimes\vctr{b}$ ($a_i b_j$). 
All inner products are indicated by a single dot between the tensorial quantities of the same order, \eg\ $\vctr{a} \cdot \vctr{b}$ ($a_i b_i$) for vectors and $\tnsr{A} \cdot \tnsr{B}$ ($A_{ij}B_{ij}$) for second-order tensors. 
The transpose, \transpose{\tnsr{A}}, of a tensor \tnsr{A} is denoted by a superscript ``T'', and the inverse, \inverse{\tnsr{A}}, by a superscript ``-1". 
Additional notation is introduced where required.
\end{document}